\pdfoutput=1
\documentclass[%
 reprint,
 amsmath,amssymb,
 aps,
 prapplied,
 superscriptaddress,
]{revtex4-2}

\usepackage{graphicx}% Include figure files
\usepackage{dcolumn}% Align table columns on decimal point
\usepackage{bm}% bold math
\begin{document}

%\preprint{APS/123-QED}

\title{Hydrogenated Amorphous Silicon Carbide: A Low-loss Deposited Dielectric for Microwave to Submillimeter Wave Superconducting Circuits}% Force line breaks with \\

\author{B. T. Buijtendorp}
\email[]{b.t.buijtendorp@tudelft.nl}
%\thanks{}
\affiliation{Faculty of Electrical Engineering, Mathematics and Computer Science, Delft University of Technology, Mekelweg 4, Delft 2628 CD, The Netherlands}

\author{S. Vollebregt}
\affiliation{Faculty of Electrical Engineering, Mathematics and Computer Science, Delft University of Technology, Mekelweg 4, Delft 2628 CD, The Netherlands}

\author{K. Karatsu}
\affiliation{Faculty of Electrical Engineering, Mathematics and Computer Science, Delft University of Technology, Mekelweg 4, Delft 2628 CD, The Netherlands}
\affiliation{SRON Netherlands Institute for Space Research, Niels Bohrweg 4, Leiden 2333 CA, The Netherlands}

\author{D. J. Thoen}
\affiliation{Faculty of Electrical Engineering, Mathematics and Computer Science, Delft University of Technology, Mekelweg 4, Delft 2628 CD, The Netherlands}

\author{V. Murugesan}
\affiliation{SRON Netherlands Institute for Space Research, Niels Bohrweg 4, Leiden 2333 CA, The Netherlands}

\author{K. Kouwenhoven}
\affiliation{Faculty of Electrical Engineering, Mathematics and Computer Science, Delft University of Technology, Mekelweg 4, Delft 2628 CD, The Netherlands}
\affiliation{SRON Netherlands Institute for Space Research, Niels Bohrweg 4, Leiden 2333 CA, The Netherlands}

\author{S. Hähnle}
\affiliation{Faculty of Electrical Engineering, Mathematics and Computer Science, Delft University of Technology, Mekelweg 4, Delft 2628 CD, The Netherlands}
\affiliation{SRON Netherlands Institute for Space Research, Niels Bohrweg 4, Leiden 2333 CA, The Netherlands}

\author{J. J. A. Baselmans}
\affiliation{Faculty of Electrical Engineering, Mathematics and Computer Science, Delft University of Technology, Mekelweg 4, Delft 2628 CD, The Netherlands}
\affiliation{SRON Netherlands Institute for Space Research, Niels Bohrweg 4, Leiden 2333 CA, The Netherlands}

\author{A. Endo}
\affiliation{Faculty of Electrical Engineering, Mathematics and Computer Science, Delft University of Technology, Mekelweg 4, Delft 2628 CD, The Netherlands}


\begin{abstract}
Low-loss deposited dielectrics will benefit superconducting devices such as integrated superconducting spectrometers, superconducting qubits and kinetic inductance parametric amplifiers. Compared with planar structures, multi-layer structures such as microstrips are more compact and eliminate radiation loss at high frequencies. Multi-layer structures are most easily fabricated with deposited dielectrics, which typically exhibit higher dielectric loss than crystalline dielectrics. We measured the sub-kelvin and low-power microwave and mm-submm wave dielectric loss of hydrogenated amorphous silicon carbide (\mbox{a-SiC:H}), using a superconducting chip with NbTiN/\mbox{a-SiC:H}/NbTiN microstrip resonators. We deposited the \mbox{a-SiC:H} by plasma-enhanced chemical vapor deposition at a substrate temperature of 400\textdegree C. The \mbox{a-SiC:H} has a mm-submm loss tangent ranging from $0.80 \pm 0.01 \times 10^{-4}$ to $1.43 \pm 0.04  \times 10^{-4}$ in the range of 270--385 GHz. The microwave loss tangent is $3.2 \pm 0.2 \times 10^{-5}$. These are the lowest low-power sub-kelvin loss tangents that have been reported for microstrip resonators at mm-submm and microwave frequencies. We observe that the loss tangent increases with frequency. The \mbox{a-SiC:H} films are free of blisters and have low stress: $-20$ MPa compressive at 200 nm thickness to 60 MPa tensile at 1000 nm thickness.
\end{abstract}

\maketitle

%\tableofcontents

Superconducting transmission lines are essential components of devices such as integrated superconducting spectrometers (ISSs) \cite{endo_first_2019, karkare_full-array_2020}, kinetic inductance parametric amplifiers (KIPAs) \cite{ho_eom_wideband_2012}, and superconducting qubits \cite{arute_quantum_2019}. These superconducting devices operate at sub-kelvin temperatures where amorphous solids exhibit excess dielectric loss compared to their crystalline counterparts due to two-level systems (TLSs) \cite{phillips_tunneling_1972, muller_towards_2019}. This is problematic since the loss limits the efficiency of ISSs \cite{hailey-dunsheath_optical_2014}, decreases the gain of KIPAs \cite{valenzuela_modelling_2019, shan_parametric_2016} and is a significant source of decoherence in superconducting qubits \cite{martinis_decoherence_2005}. The problem is often circumvented by avoiding deposited dielectrics, for example by employing planar structures. However, a low-loss deposited dielectric will enable the use of microstrips, thereby achieving benefits such as miniaturization and the removal of radiation loss \cite{hahnle_suppression_2020}. 

With hydrogenated amorphous silicon (\mbox{a-Si:H}) in microstrip resonators, sub-kelvin and low-power loss tangents $\tan{\delta}$ of 4--5$\times 10^{-5}$ have been reported at microwave frequencies \cite{mazin_thin_2010, oconnell_microwave_2008, hahnle_superconducting_2021, buijtendorp_characterization_2020}.  Recently for a-Si:H a $\tan{\delta}$ of $2.1 \pm 0.1  \times 10^{-4}$ has been reported at 350 GHz \cite{hahnle_superconducting_2021}. Although this is an improvement over $\mathrm{SiO_2}$ \cite{oconnell_microwave_2008,gao_measurement_2009} and $\mathrm{SiN}_x$ \cite{oconnell_microwave_2008, endo_-chip_2013, hailey-dunsheath_optical_2014}, a further reduction in $\tan{\delta}$ is desirable. For example, the efficiency of the mm-submm filters of an ISS depends on $R \cdot \tan\delta$ \cite{hailey-dunsheath_optical_2014}, where $R = f / \Delta f$ is the resolving power. With the current mm-submm $\tan{\delta}$ of $2.1 \times 10^{-4}$ of a-Si:H the filters are limited to $R \sim 500$. A higher $R$ requires a reduction of $\tan{\delta}$. 

Although the microscopic origin of the TLSs remains unknown \cite{phillips_tunneling_1972, muller_towards_2019}, a comparison of $\mathrm{SiO_2}$, $\mathrm{SiN}_x$, \mbox{a-Si:H} and crystalline Si (c-Si) has led to the hypothesis that a more constrained lattice (i.e. an increase in coordination number) is correlated with a decrease in dielectric loss \cite{oconnell_microwave_2008}. With electron-beam evaporated amorphous silicon (a-Si) it has been observed that a lower microwave loss can be achieved by depositing at elevated substrate temperatures ($T_\mathrm{sub}$), and a correlation was found between the microwave loss and the dangling bond density \cite{molina-ruiz_origin_2021}. The electron-beam evaporated and sputtered a-Si films, which are hydrogen-free, exhibit an order of magnitude larger microwave loss than the \mbox{a-Si:H} \cite{molina-ruiz_origin_2021,oconnell_microwave_2008}, suggesting that there is a relation between non-passivated dangling bonds and dielectric loss. 

Similar to \mbox{a-Si:H}, \mbox{a-SiC:H} is a hydrogenated and four-fold coordinated material. Deposition of low-stress \mbox{a-SiC:H} films at high $T_\mathrm{sub}$ is possible with plasma-enhanced chemical vapor deposition (PECVD) \cite{sarro_low-stress_1998}. The \mbox{a-SiC:H} has high wet etching selectivity with Si \cite{sarro_low-stress_1998}, enabling an \mbox{a-SiC:H} film to be used as a membrane, for example in superconducting photon detectors \cite{de_visser_phonon-trapping_2021}. The \mbox{a-SiC:H} can be patterned by reactive ion etching (RIE). Although the dielectric constants of a-SiC:H and of a-Si:H depend on the deposition parameters, a-SiC:H generally has a lower microwave $\epsilon_\mathrm{r} \approx 7$ than the \mbox{a-Si:H} microwave $\epsilon_\mathrm{r} \approx 10$. The \mbox{a-SiC:H} films do not exhibit blisters, which are a common issue with \mbox{a-Si:H} \cite{mishima_investigation_1988, wang_avoiding_2018}. These properties suggest that \mbox{a-SiC:H} is a promising alternative to \mbox{a-Si:H} and $\mathrm{SiN}_x$ for superconducting devices. However, the sub-kelvin and low-power microwave and mm-submm wave dielectric loss of \mbox{a-SiC:H} have not yet been reported.

Here we present measurements of the sub-kelvin and low-power dielectric loss of \mbox{a-SiC:H} at microwave (7~GHz) and mm-submm (270--385~GHz) frequencies, using a lab-on-chip device with superconducting NbTiN/\mbox{a-SiC:H}/NbTiN microstrip resonators. The fabrication details for an equivalent chip with a-Si:H instead of a-SiC:H have previously been reported \cite{hahnle_superconducting_2021}. The 295 nm thick \mbox{a-SiC:H} film was deposited by PECVD using a Novellus Concept One \cite{van_de_ven_advances_1988}. The precursor gasses were silane ($\mathrm{SiH_4}$) and methane ($\mathrm{CH_4}$) with flow rates of 25~sccm and 411~sccm respectively, and no dilution gas. We deposited with a $T_\mathrm{sub}$ of 400\textdegree C, chamber pressure of 2~Torr, 450 kHz RF power of 150~W, and 13.56 MHz RF power of 450~W. The \mbox{a-SiC:H} films exhibit low stress, ranging from $-20$ MPa compressive at 200 nm thickness to 60~MPa tensile at 1000~nm thickness. We patterned the \mbox{a-SiC:H} layer using RIE with an $\mathrm{SF_6}$ and $\mathrm{O_2}$ plasma at a rate of approximately 0.8~nm/s. 

\begin{figure}
\includegraphics[width=\linewidth]{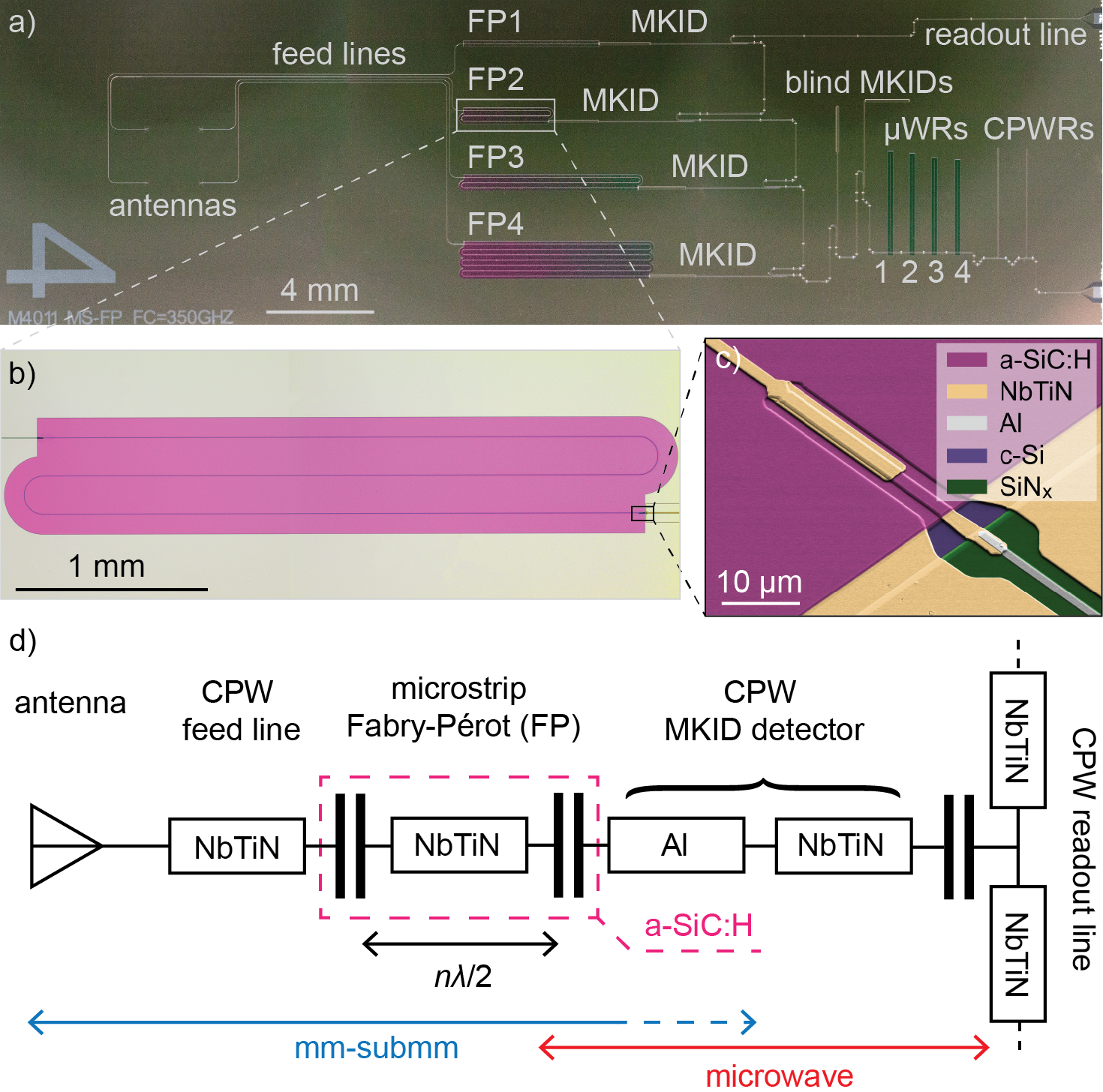}% Here is how to import EPS art
\caption{\label{fig:SEM} a) Photograph of the chip. The four Fabry-Pérot resonators (FP1--4) receive the mm-submm signal from antennas that are coupled to a photomixer source. Each FP is coupled to a microwave kinetic inductance detector (MKID). The four shunted microstrip microwave resonators ($\mathrm{\mu}$WR1--4) are coupled to the readout line. b) Micrograph of Fabry-Pérot resonator 2 (FP2). The purple area is \mbox{a-SiC:H}. The feed line is visible on the left of the image and the MKID is visible on the right. c)  Tilted scanning electron micrograph of the coupling of FP2 to the MKID, with false coloring. d) Schematic of an antenna--FP--MKID circuit. The FPs have a length of an integer (mode number $n$) multiple  of half the mm-submm wavelength $\lambda$, and have transmission peaks at $n f_\mathrm{0}$, where $f_\mathrm{0}$ is the fundamental resonance frequency. The mm-submm signal is absorbed in the Al section of the MKID, resulting in a change of resonance frequency (frequency response) of the MKID.}
\end{figure}

\begin{figure*}
\includegraphics[width=\textwidth]{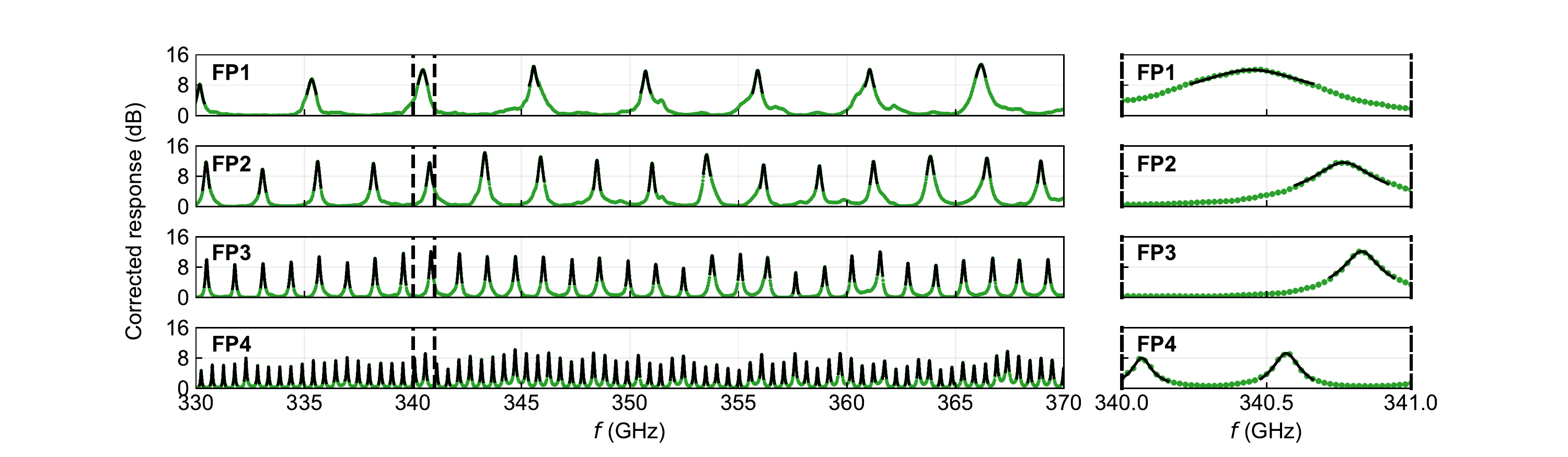}%Fig3 Here is how to import EPS art
\caption{\label{fig:peaks}The measured frequency response in the 330--370 GHz band. The response is measured by the MKIDs that are coupled to the FP resonators 1--4. The response is corrected for stray light by dividing by the response of an MKID that is not coupled to an FP resonator (blind MKID in Fig. 1a). The green dots represent the measurement data, and the black curves are Lorentzian fits to the peaks, from which we obtain the average $Q$ for each FP resonator. The same analysis is performed for each of the four frequency bands. The right plot shows the same data zoomed in to 340--341 GHz.}
\end{figure*}

Fig. 1a shows a photograph of the device that we used to measure the dielectric loss. The device and experimental setup are equivalent to what has been reported for earlier work on a-Si:H \cite{hahnle_superconducting_2021}. To measure the mm-submm loss we analyzed the transmission through four in-line Fabry-Pérot NbTiN/\mbox{a-SiC:H}/NbTiN microstrip resonators (FP1--4) with 2 $\mathrm{\mu m}$ line width. Each FP is coupled to a NbTiN CPW feed line on one end, and a hybrid NbTiN-Al microwave kinetic inductance detector (MKID) \cite{janssen_high_2013} on the other end. Each feed line is connected to a twin slot antenna with a center frequency of 350 GHz located in the focus of a Si-lens glued to the backside of the chip, which receives mm-submm radiation from a photomixer source. The MKIDs are coupled to a shared NbTiN CPW microwave readout line. In Fig. 1b we show a micrograph of FP2, where the FP is coupled to the feed line on the left side, and coupled to the MKID on the right side. In Fig. 1c we show a scanning electron micrograph of the coupling of FP2 to an MKID. 

A schematic of the mm-submm experiment is given in Fig. 1d. Each FP has a different length, resulting in different fundamental resonance frequencies $f_\mathrm{0}$. The FPs have transmission peaks at integer multiples (mode numbers $n$) of $f_\mathrm{0}$. The power that is transmitted through an FP resonator breaks Cooper pairs in the Al section of the MKID, resulting in a change in the MKID's resonance frequency. The readout of the MKIDs is performed using a frequency-multiplexed readout system \cite{van_rantwijk_multiplexed_2016}. The frequency responses trace the FPs' transmission curves $|S_\mathrm{21}(f)|^2$ and therefore it can be used to obtain the FPs' loaded quality factors $Q$ \cite{hahnle_superconducting_2021}. The frequency response was corrected for stray light by dividing by the response of an MKID that is not connected to an FP resonator (blind MKID in Fig. 1a). The measurements were performed at 120 mK. We estimate that the power absorbed by the MKIDs is on the order of a few picowatts, corresponding to single photon energies inside the FP resonators. We observe no significant correlation between $Q_\mathrm{i}$ and the internal power inside the FP resonators. We confirmed that the MKIDs exhibit a linear response by observing that there is no significant correlation between the measured $Q$ and the power absorbed by the MKIDs. 

From the standard tunneling model (STM) \cite{phillips_tunneling_1972} for TLSs it follows that the loss tangent $\tan \delta$ is dependent on the temperature $T$, frequency $f$ and average number of photons in the resonator $N$ \cite{gao_physics_2008}: 
\begin{equation}\label{eq:tand}
\tan \delta = \tan \delta_\mathrm{0} \tanh \frac{h f}{2 k_B T} \left(1 + \frac{N}{N_\mathrm{0}}\right)^{-\beta /2} + \tan \delta_\mathrm{HP}
,\end{equation}
where $\tan \delta_\mathrm{0}$ is the TLS-induced loss tangent in the low-power limit and at 0 K, $N_\mathrm{0}$ is the critical photon number above which the TLSs saturate, and $\beta$ is equal to 1 in the STM, but it has previously been observed in the range of 0.3--0.7 \cite{molina-ruiz_origin_2021}. The $\tan \delta_\mathrm{HP}$ term represents losses other than TLS loss that dominate at high internal resonator power. The $\tan{\delta}$ is obtained from the resonator internal quality factor $Q_\mathrm{i}$:
\begin{equation}
\tan{\delta} = 1/(pQ_\mathrm{i})
,\end{equation}
where $p$ is the filling fraction \cite{gao_physics_2008} of the dielectric, which we determined to be 0.97 in our microstrip resonators using the EM-field solver Sonnet \cite{sonnet_user_guide}.

\begin{figure}
\includegraphics[width=\linewidth]{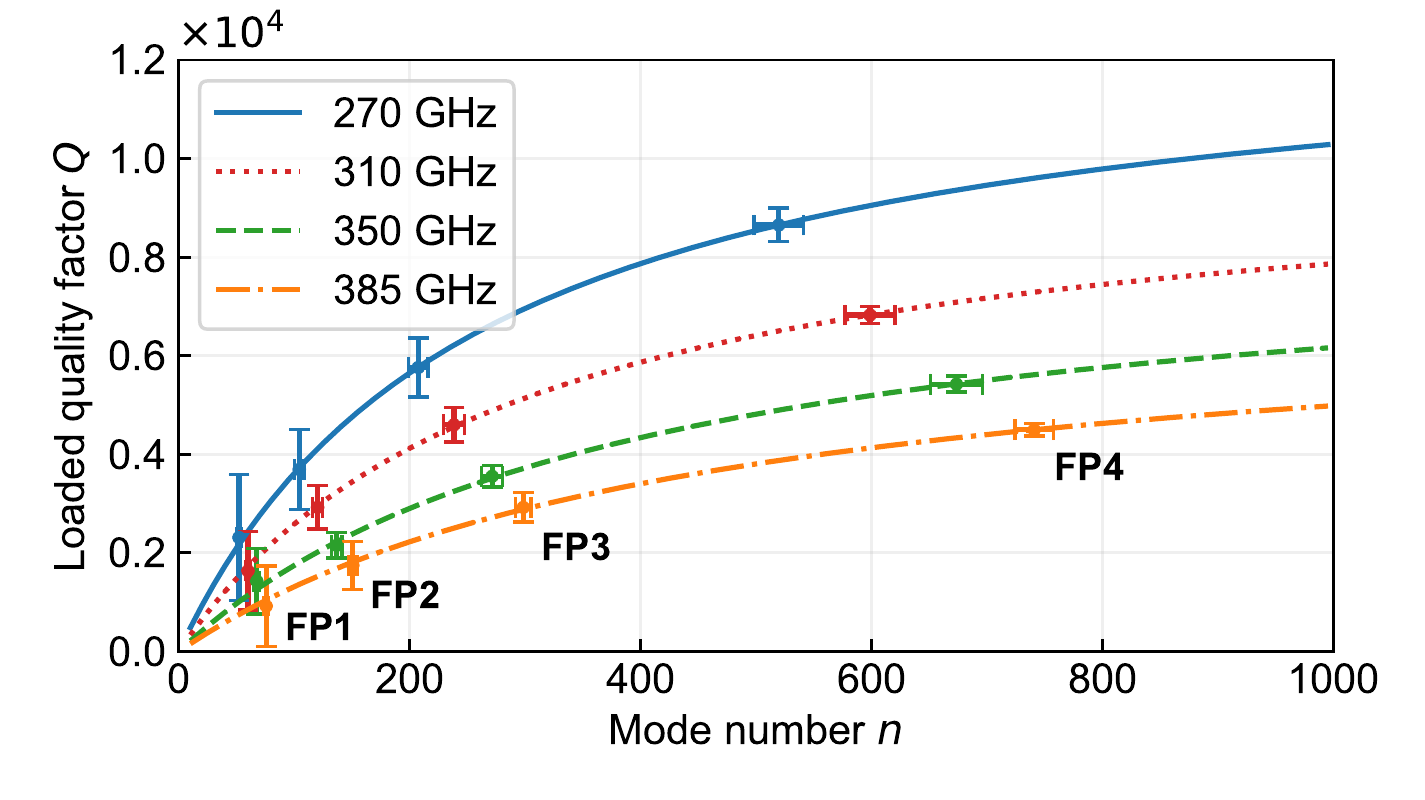}% Here is how to import EPS art
\caption{\label{fig:2} Loaded quality factor $Q$ versus mode number $n$ of the Fabry-Pérot resonators (FP1--4), for the frequencies shown in the legend. The curves are fits of Eq. \ref{eq:Ql_n}. The vertical error bars represent 15 standard errors ($\pm 15 \, \mathrm{SE}$) in $Q$, and the $1/SE$ were used as fitting weights. The horizontal error bars represent one standard deviation ($\pm \sigma$) in $n$.}
\end{figure}

\begin{figure}
\includegraphics[width=\linewidth]{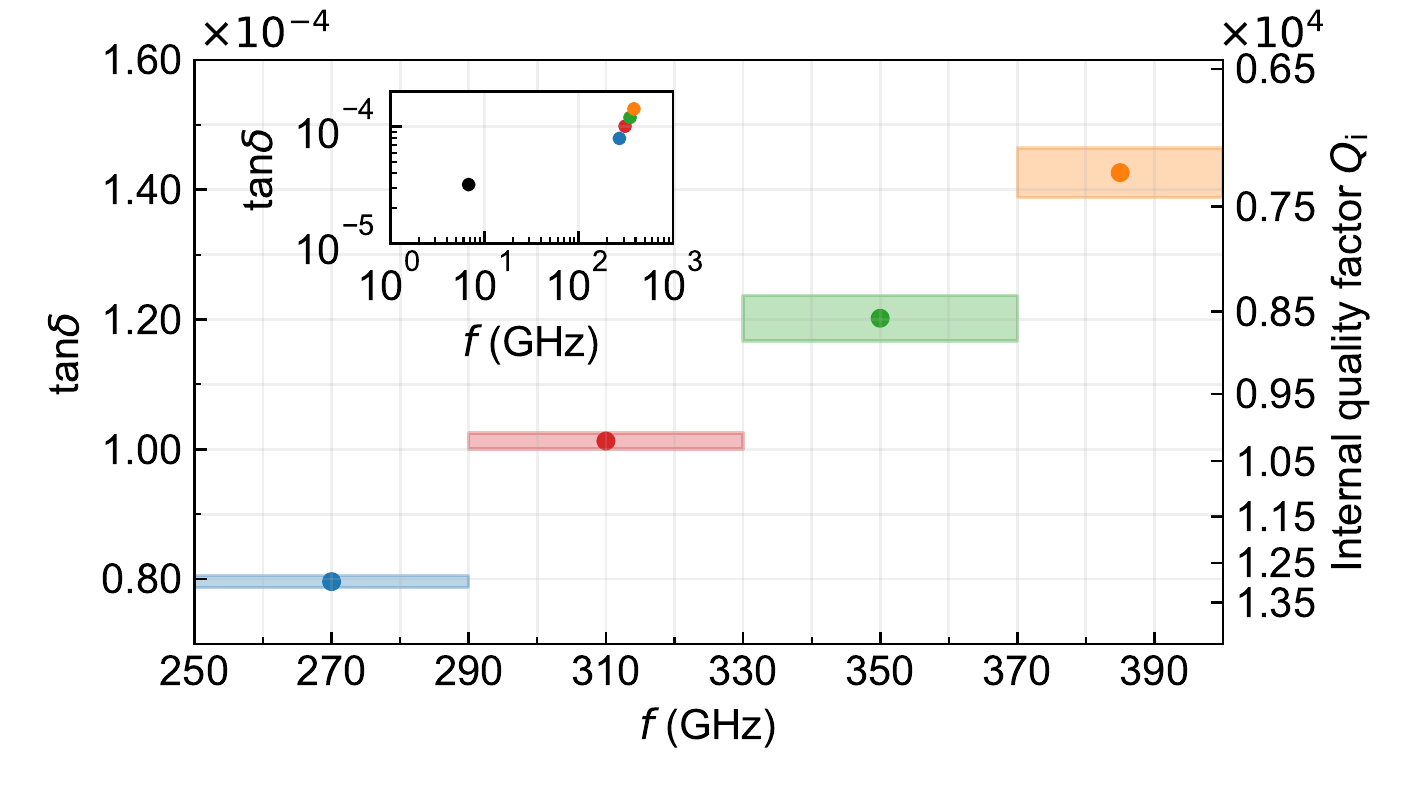}% Here is how to import EPS art
\caption{\label{fig:4} The mm-submm $\tan \delta = 1/\left( p Q_\mathrm{i} \right)$ versus frequency, which we obtained by fitting Eq. \ref{eq:Ql_n} to the data in Fig. \ref{fig:2}. The dots represent the measured $\tan{\delta}$ values and the shaded areas represent one standard deviation ($\pm \sigma$) uncertainty in $\tan{\delta}$ in the vertical direction, and the measured frequency range over which $Q$ and $n$ were averaged in the horizontal direction. The inset also shows the average microwave $\tan{\delta_0}$ of $\mathrm{\mu}$WR1--4. We plot the microwave data in Fig. \ref{fig:1}.}
\end{figure}

To determine $\tan 
\delta$ at 270, 310, 350 and 385~GHz we obtained the $Q_\mathrm{i}$ at each frequency by fitting the measured $Q$ versus $n$ data of the FP resonators to the equation \cite{hahnle_superconducting_2021}
\begin{equation}\label{eq:Ql_n}
    Q = \frac{nQ_\mathrm{c,1}Q_\mathrm{i}}{nQ_\mathrm{c,1} + Q_\mathrm{i}}
,\end{equation}
where $Q_\mathrm{c,1} = Q_\mathrm{c}/n$, with $Q_\mathrm{c}$ the coupling quality factor. The fits to Eq. \ref{eq:Ql_n} are plotted in Fig. \ref{fig:2}. We obtained the average $n$ and the average $Q$ from Lorentzian fits to the FP peaks in four frequency bands centered around these frequencies. The ranges of the frequency bands are visible in Fig. \ref{fig:4}. We show the Lorentzian fits in the 330--370 GHz band in Fig. \ref{fig:peaks}. We plot the resulting $\tan{\delta}$ for each frequency in Fig. \ref{fig:4}. 

We observe that the $\tan{\delta}$ is frequency dependent: It ranges from $0.80 \pm 0.01 \times 10^{-4}$ to $1.43 \pm 0.04  \times 10^{-4}$ in the range of 270--385 GHz. These $\tan{\delta}$ values are lower than what has been reported for other dielectrics such as \mbox{a-Si:H} \cite{hahnle_superconducting_2021} and $\mathrm{SiN}_x$ \cite{endo_-chip_2013, hailey-dunsheath_optical_2014}. At 350 GHz the \mbox{a-SiC:H} exhibits a $\tan{\delta}$ of $1.20 \pm 0.04  \times 10^{-4}$, compared with a $\tan{\delta}$ of $2.1 \pm 0.1  \times 10^{-4}$ which was measured for \mbox{a-Si:H} using the same experimental method \cite{hahnle_superconducting_2021}. 

\begin{figure}
\includegraphics[width=\linewidth]{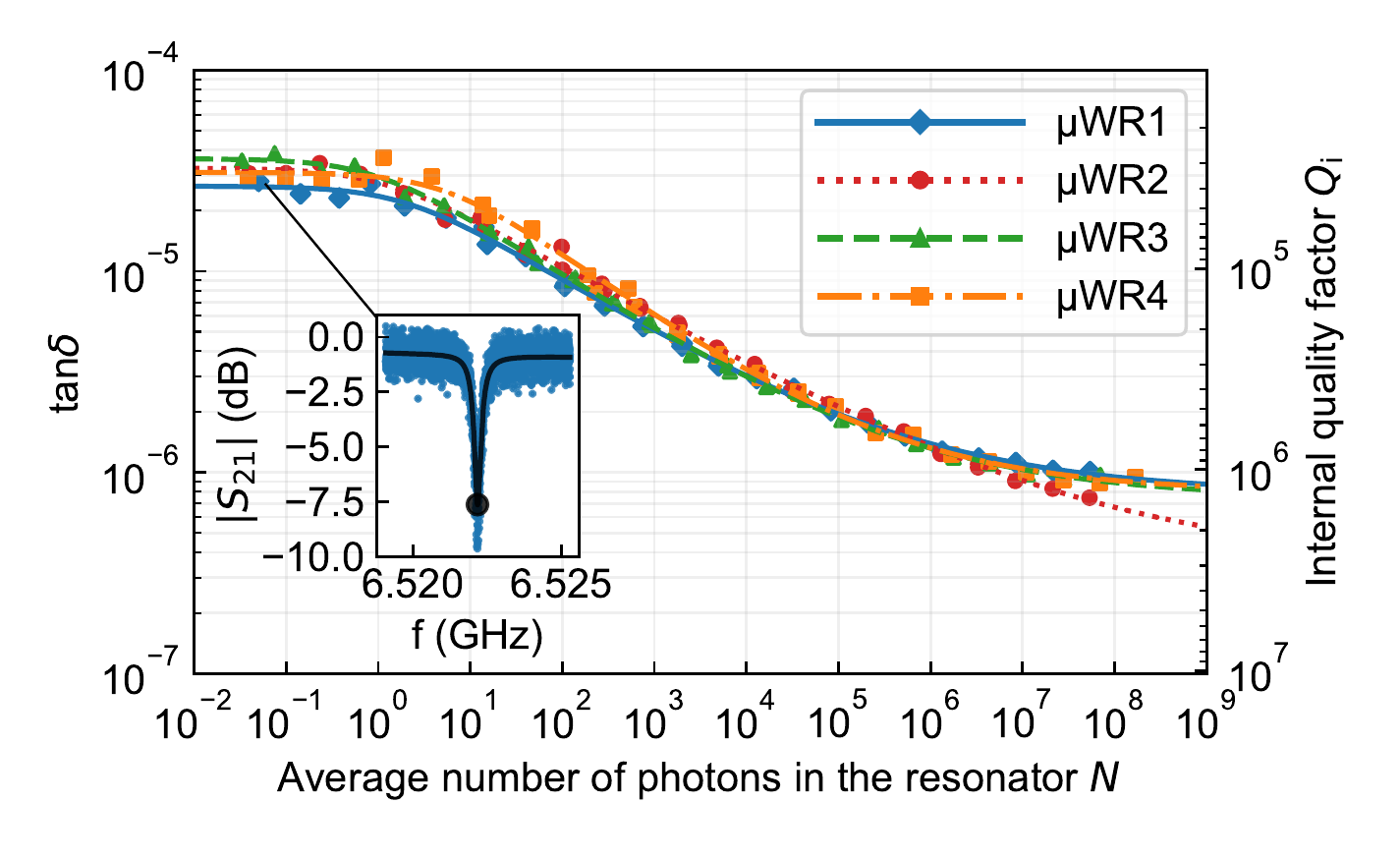}% Here is how to import EPS art
\caption{\label{fig:1} Measured microwave $\tan \delta = 1/\left( p Q_\mathrm{i} \right)$ versus average number of photons in the resonator, of $\mathrm{\mu}$WR1--4. The curves are fits to Eq. \ref{eq:tand}, resulting in an average $\tan \delta_\mathrm{0}$ of $3.2 \pm 0.2 \times 10^{-5}$. The inset illustrates how $\tan \delta$ is obtained at each power, by fitting the $|S_\mathrm{21}|$ dips using the model in Ref. \cite{khalil_analysis_2012}.}
\end{figure}

In addition to measuring the mm-submm $\tan \delta$, we measured the microwave $\tan \delta$ of the a-SiC:H using four NbTiN/a-SiC:H/NbTiN half-wavelength microstrip microwave resonators ($\mathrm{\mu}$WR1--4 in Fig. 1a.) with 2 $\mathrm{\mu m}$ line width. We used a vector network analyzer to measure the $S_{21}$-parameter of the NbTiN coplanar waveguide (CPW) readout line, which is coupled to the four shunted $\mathrm{\mu}$WRs. We measured $\tan \delta$ in a range of internal resonator powers, corresponding to an average number of photons per half-wavelength in the resonators $N$  of roughly $10^{-1}$--$10^8$. $N$ is equal to
\begin{equation}
    N = \frac{P_\mathrm{int}}{hf^2}
,\end{equation}
where $P_\mathrm{int}$ is the resonator's internal power. We fitted the $|S_\mathrm{21}|$ dips of the four $\mathrm{\mu WRs}$ using the model described in Ref. \cite{khalil_analysis_2012}, to obtain the resonators' $Q_\mathrm{i}$. For illustration we show one of the $|S_\mathrm{21}|$ dips in the inset of Fig. \ref{fig:1}. 

We obtained $\tan \delta_\mathrm{0}$ and $\beta$ by fitting Eq. \ref{eq:tand} to the measured $\tan\delta$ versus $N$, as shown in Fig. \ref{fig:1}. The $\mathrm{\mu WRs}$ exhibit an average $\tan \delta_\mathrm{0}$ of $3.2 \pm 0.2 \times 10^{-5}$. This is comparable to what has been reported for \mbox{a-Si:H} \cite{oconnell_microwave_2008,mazin_thin_2010,hahnle_superconducting_2021}. We measured $\beta$ values in the range of 0.5--0.7. We confirm that the low-power loss is dominated by the \mbox{a-SiC:H} by measuring a shunted NbTiN CPW quarter-wavelength resonator (CPWR) that was fabricated directly on top of the c-Si substrate, which exhibits a $\tan \delta_0$ of $4.1 \pm 0.1 \times 10^{-6}$.

For the a-SiC:H we derive a sub-kelvin dielectric constant $\varepsilon_\mathrm{r} \approx 7.0$ for microwaves and mm-submm waves, based on the resonance frequencies of the FP resonators and the $\mathrm{\mu}$WRs. We measured a room temperature $\varepsilon_\mathrm{r}$ of 6.9 at 15--66 THz using Fourier-transform infrared spectroscopy and an $\varepsilon_\mathrm{r}$ of 7.5 at 250 THz (1200 nm) using ellipsometry. We observe a bandgap of 1.8 eV, which we obtained by fitting the Tauc-Lorentz dispersion model to the ellipsometry data in the range of 0.8--2.7 eV.

The observed frequency dependence of the dielectric loss of \mbox{a-SiC:H}, throughout microwave to mm-submm frequencies, is not explained by the STM which assumes a frequency independent TLS density \cite{phillips_tunneling_1972}. A frequency-dependent mm-submm $\tan \delta$ has previously been observed for $\mathrm{SiN}_x$ \cite{endo_-chip_2013}, and a larger mm-submm $\tan \delta$ than microwave $\tan \delta$ has been observed for both $\mathrm{SiN}_x$ \cite{endo_-chip_2013,oconnell_microwave_2008} and \mbox{a-Si:H} \cite{hahnle_superconducting_2021,oconnell_microwave_2008}. For a-Si:H \cite{hahnle_superconducting_2021} it has been suggested that the mm-submm loss can be caused by phonon modes at far-infrared wavelengths \cite{cataldo_infrared_2012, cataldo_infrared_2016}, and the same could be true for a-SiC:H. Alternatively, a generalized TLS model could account for a frequency dependent $\tan \delta$ \cite{faoro_interacting_2015}. If the mm-submm loss of \mbox{a-SiC:H} is dominated by TLS loss, it is still an open question if the loss is mainly due to the bulk of the dielectric, or due to TLSs in the interface layers \cite{woods_determining_2019, gao_experimental_2008}. 

To conclude, PECVD \mbox{a-SiC:H} exhibits the lowest low-power sub-kelvin mm-submm $\tan \delta$ that has been reported to date for microstrip resonators. The microwave $\tan \delta$ is comparable to the best values that have been reported for microstrip resonators \cite{hahnle_superconducting_2021, mazin_thin_2010}. Interestingly, we observe a frequency dependence of $\tan \delta$ which is not explained by the STM. In addition to the low $\tan \delta$, the \mbox{a-SiC:H} films are free of blisters and exhibit low stress. These properties make \mbox{a-SiC:H} a promising dielectric for microwave to submm wave superconducting circuits in many applications.

\section*{Acknowledgements}

We thank the staff of the Else Kooi Laboratory and the Kavli Nanolab Delft for their support. A.E. was supported by the Netherlands Organization for Scientific Research NWO (Vidi grant no. 639.042.423). The contribution of J.J.A. Baselmans was supported by the ERC CoG 648135 MOSAIC.

\bibliography{ms.bib}% Produces the bibliography via BibTeX.

\providecommand{\noopsort}[1]{}\providecommand{\singleletter}[1]{#1}%
\begin{thebibliography}{33}%
\makeatletter
\providecommand \@ifxundefined [1]{%
 \@ifx{#1\undefined}
}%
\providecommand \@ifnum [1]{%
 \ifnum #1\expandafter \@firstoftwo
 \else \expandafter \@secondoftwo
 \fi
}%
\providecommand \@ifx [1]{%
 \ifx #1\expandafter \@firstoftwo
 \else \expandafter \@secondoftwo
 \fi
}%
\providecommand \natexlab [1]{#1}%
\providecommand \enquote  [1]{``#1''}%
\providecommand \bibnamefont  [1]{#1}%
\providecommand \bibfnamefont [1]{#1}%
\providecommand \citenamefont [1]{#1}%
\providecommand \href@noop [0]{\@secondoftwo}%
\providecommand \href [0]{\begingroup \@sanitize@url \@href}%
\providecommand \@href[1]{\@@startlink{#1}\@@href}%
\providecommand \@@href[1]{\endgroup#1\@@endlink}%
\providecommand \@sanitize@url [0]{\catcode `\\12\catcode `\$12\catcode
  `\&12\catcode `\#12\catcode `\^12\catcode `\_12\catcode `\%12\relax}%
\providecommand \@@startlink[1]{}%
\providecommand \@@endlink[0]{}%
\providecommand \url  [0]{\begingroup\@sanitize@url \@url }%
\providecommand \@url [1]{\endgroup\@href {#1}{\urlprefix }}%
\providecommand \urlprefix  [0]{URL }%
\providecommand \Eprint [0]{\href }%
\providecommand \doibase [0]{https://doi.org/}%
\providecommand \selectlanguage [0]{\@gobble}%
\providecommand \bibinfo  [0]{\@secondoftwo}%
\providecommand \bibfield  [0]{\@secondoftwo}%
\providecommand \translation [1]{[#1]}%
\providecommand \BibitemOpen [0]{}%
\providecommand \bibitemStop [0]{}%
\providecommand \bibitemNoStop [0]{.\EOS\space}%
\providecommand \EOS [0]{\spacefactor3000\relax}%
\providecommand \BibitemShut  [1]{\csname bibitem#1\endcsname}%
\let\auto@bib@innerbib\@empty
%</preamble>
\bibitem [{\citenamefont {Endo}\ \emph {et~al.}(2019)\citenamefont {Endo},
  \citenamefont {Karatsu}, \citenamefont {Tamura}, \citenamefont {Oshima},
  \citenamefont {Taniguchi}, \citenamefont {Takekoshi}, \citenamefont
  {Asayama}, \citenamefont {Bakx}, \citenamefont {Bosma}, \citenamefont
  {Bueno}, \citenamefont {Chin}, \citenamefont {Fujii}, \citenamefont {Fujita},
  \citenamefont {Huiting}, \citenamefont {Ikarashi}, \citenamefont {Ishida},
  \citenamefont {Ishii}, \citenamefont {Kawabe}, \citenamefont {Klapwijk},
  \citenamefont {Kohno}, \citenamefont {Kouchi}, \citenamefont {Llombart},
  \citenamefont {Maekawa}, \citenamefont {Murugesan}, \citenamefont
  {Nakatsubo}, \citenamefont {Naruse}, \citenamefont {Ohtawara}, \citenamefont
  {Pascual~Laguna}, \citenamefont {Suzuki}, \citenamefont {Suzuki},
  \citenamefont {Thoen}, \citenamefont {Tsukagoshi}, \citenamefont {Ueda},
  \citenamefont {de~Visser}, \citenamefont {van~der Werf}, \citenamefont
  {Yates}, \citenamefont {Yoshimura}, \citenamefont {Yurduseven},\ and\
  \citenamefont {Baselmans}}]{endo_first_2019}%
  \BibitemOpen
  \bibfield  {author} {\bibinfo {author} {\bibfnamefont {A.}~\bibnamefont
  {Endo}}, \bibinfo {author} {\bibfnamefont {K.}~\bibnamefont {Karatsu}},
  \bibinfo {author} {\bibfnamefont {Y.}~\bibnamefont {Tamura}}, \bibinfo
  {author} {\bibfnamefont {T.}~\bibnamefont {Oshima}}, \bibinfo {author}
  {\bibfnamefont {A.}~\bibnamefont {Taniguchi}}, \bibinfo {author}
  {\bibfnamefont {T.}~\bibnamefont {Takekoshi}}, \bibinfo {author}
  {\bibfnamefont {S.}~\bibnamefont {Asayama}}, \bibinfo {author} {\bibfnamefont
  {T.~J. L.~C.}\ \bibnamefont {Bakx}}, \bibinfo {author} {\bibfnamefont
  {S.}~\bibnamefont {Bosma}}, \bibinfo {author} {\bibfnamefont
  {J.}~\bibnamefont {Bueno}}, \bibinfo {author} {\bibfnamefont {K.~W.}\
  \bibnamefont {Chin}}, \bibinfo {author} {\bibfnamefont {Y.}~\bibnamefont
  {Fujii}}, \bibinfo {author} {\bibfnamefont {K.}~\bibnamefont {Fujita}},
  \bibinfo {author} {\bibfnamefont {R.}~\bibnamefont {Huiting}}, \bibinfo
  {author} {\bibfnamefont {S.}~\bibnamefont {Ikarashi}}, \bibinfo {author}
  {\bibfnamefont {T.}~\bibnamefont {Ishida}}, \bibinfo {author} {\bibfnamefont
  {S.}~\bibnamefont {Ishii}}, \bibinfo {author} {\bibfnamefont
  {R.}~\bibnamefont {Kawabe}}, \bibinfo {author} {\bibfnamefont {T.~M.}\
  \bibnamefont {Klapwijk}}, \bibinfo {author} {\bibfnamefont {K.}~\bibnamefont
  {Kohno}}, \bibinfo {author} {\bibfnamefont {A.}~\bibnamefont {Kouchi}},
  \bibinfo {author} {\bibfnamefont {N.}~\bibnamefont {Llombart}}, \bibinfo
  {author} {\bibfnamefont {J.}~\bibnamefont {Maekawa}}, \bibinfo {author}
  {\bibfnamefont {V.}~\bibnamefont {Murugesan}}, \bibinfo {author}
  {\bibfnamefont {S.}~\bibnamefont {Nakatsubo}}, \bibinfo {author}
  {\bibfnamefont {M.}~\bibnamefont {Naruse}}, \bibinfo {author} {\bibfnamefont
  {K.}~\bibnamefont {Ohtawara}}, \bibinfo {author} {\bibfnamefont
  {A.}~\bibnamefont {Pascual~Laguna}}, \bibinfo {author} {\bibfnamefont
  {J.}~\bibnamefont {Suzuki}}, \bibinfo {author} {\bibfnamefont
  {K.}~\bibnamefont {Suzuki}}, \bibinfo {author} {\bibfnamefont {D.~J.}\
  \bibnamefont {Thoen}}, \bibinfo {author} {\bibfnamefont {T.}~\bibnamefont
  {Tsukagoshi}}, \bibinfo {author} {\bibfnamefont {T.}~\bibnamefont {Ueda}},
  \bibinfo {author} {\bibfnamefont {P.~J.}\ \bibnamefont {de~Visser}}, \bibinfo
  {author} {\bibfnamefont {P.~P.}\ \bibnamefont {van~der Werf}}, \bibinfo
  {author} {\bibfnamefont {S.~J.~C.}\ \bibnamefont {Yates}}, \bibinfo {author}
  {\bibfnamefont {Y.}~\bibnamefont {Yoshimura}}, \bibinfo {author}
  {\bibfnamefont {O.}~\bibnamefont {Yurduseven}},\ and\ \bibinfo {author}
  {\bibfnamefont {J.~J.~A.}\ \bibnamefont {Baselmans}},\ }\bibfield  {title}
  {\bibinfo {title} {First light demonstration of the integrated
  superconducting spectrometer},\ }\href@noop {} {\bibfield  {journal}
  {\bibinfo  {journal} {Nat. Astron.}\ }\textbf {\bibinfo {volume} {3}},\
  \bibinfo {pages} {989} (\bibinfo {year} {2019})}\BibitemShut {NoStop}%
\bibitem [{\citenamefont {Karkare}\ \emph {et~al.}(2020)\citenamefont
  {Karkare}, \citenamefont {Barry}, \citenamefont {Bradford}, \citenamefont
  {Chapman}, \citenamefont {Doyle}, \citenamefont {Glenn}, \citenamefont
  {Gordon}, \citenamefont {Hailey-Dunsheath}, \citenamefont {Janssen},
  \citenamefont {Kovács}, \citenamefont {LeDuc}, \citenamefont {Mauskopf},
  \citenamefont {McGeehan}, \citenamefont {Redford}, \citenamefont {Shirokoff},
  \citenamefont {Tucker}, \citenamefont {Wheeler},\ and\ \citenamefont
  {Zmuidzinas}}]{karkare_full-array_2020}%
  \BibitemOpen
  \bibfield  {author} {\bibinfo {author} {\bibfnamefont {K.~S.}\ \bibnamefont
  {Karkare}}, \bibinfo {author} {\bibfnamefont {P.~S.}\ \bibnamefont {Barry}},
  \bibinfo {author} {\bibfnamefont {C.~M.}\ \bibnamefont {Bradford}}, \bibinfo
  {author} {\bibfnamefont {S.}~\bibnamefont {Chapman}}, \bibinfo {author}
  {\bibfnamefont {S.}~\bibnamefont {Doyle}}, \bibinfo {author} {\bibfnamefont
  {J.}~\bibnamefont {Glenn}}, \bibinfo {author} {\bibfnamefont
  {S.}~\bibnamefont {Gordon}}, \bibinfo {author} {\bibfnamefont
  {S.}~\bibnamefont {Hailey-Dunsheath}}, \bibinfo {author} {\bibfnamefont
  {R.~M.~J.}\ \bibnamefont {Janssen}}, \bibinfo {author} {\bibfnamefont
  {A.}~\bibnamefont {Kovács}}, \bibinfo {author} {\bibfnamefont {H.~G.}\
  \bibnamefont {LeDuc}}, \bibinfo {author} {\bibfnamefont {P.}~\bibnamefont
  {Mauskopf}}, \bibinfo {author} {\bibfnamefont {R.}~\bibnamefont {McGeehan}},
  \bibinfo {author} {\bibfnamefont {J.}~\bibnamefont {Redford}}, \bibinfo
  {author} {\bibfnamefont {E.}~\bibnamefont {Shirokoff}}, \bibinfo {author}
  {\bibfnamefont {C.}~\bibnamefont {Tucker}}, \bibinfo {author} {\bibfnamefont
  {J.}~\bibnamefont {Wheeler}},\ and\ \bibinfo {author} {\bibfnamefont
  {J.}~\bibnamefont {Zmuidzinas}},\ }\bibfield  {title} {\bibinfo {title}
  {Full-{Array} {Noise} {Performance} of {Deployment}-{Grade} {SuperSpec}
  mm-{Wave} {On}-{Chip} {Spectrometers}},\ }\href@noop {} {\bibfield  {journal}
  {\bibinfo  {journal} {J. Low Temp. Phys.}\ }\textbf {\bibinfo {volume}
  {199}},\ \bibinfo {pages} {849} (\bibinfo {year} {2020})}\BibitemShut
  {NoStop}%
\bibitem [{\citenamefont {Ho~Eom}\ \emph {et~al.}(2012)\citenamefont {Ho~Eom},
  \citenamefont {Day}, \citenamefont {LeDuc},\ and\ \citenamefont
  {Zmuidzinas}}]{ho_eom_wideband_2012}%
  \BibitemOpen
  \bibfield  {author} {\bibinfo {author} {\bibfnamefont {B.}~\bibnamefont
  {Ho~Eom}}, \bibinfo {author} {\bibfnamefont {P.~K.}\ \bibnamefont {Day}},
  \bibinfo {author} {\bibfnamefont {H.~G.}\ \bibnamefont {LeDuc}},\ and\
  \bibinfo {author} {\bibfnamefont {J.}~\bibnamefont {Zmuidzinas}},\ }\bibfield
   {title} {\bibinfo {title} {A wideband, low-noise superconducting amplifier
  with high dynamic range},\ }\href@noop {} {\bibfield  {journal} {\bibinfo
  {journal} {Nat. Phys.}\ }\textbf {\bibinfo {volume} {8}},\ \bibinfo {pages}
  {623} (\bibinfo {year} {2012})}\BibitemShut {NoStop}%
\bibitem [{\citenamefont {Arute}\ \emph {et~al.}(2019)\citenamefont {Arute},
  \citenamefont {Arya}, \citenamefont {Babbush}, \citenamefont {Bacon},
  \citenamefont {Bardin}, \citenamefont {Barends}, \citenamefont {Biswas},
  \citenamefont {Boixo}, \citenamefont {Brandao}, \citenamefont {Buell},
  \citenamefont {Burkett}, \citenamefont {Chen}, \citenamefont {Chen},
  \citenamefont {Chiaro}, \citenamefont {Collins}, \citenamefont {Courtney},
  \citenamefont {Dunsworth}, \citenamefont {Farhi}, \citenamefont {Foxen},
  \citenamefont {Fowler}, \citenamefont {Gidney}, \citenamefont {Giustina},
  \citenamefont {Graff}, \citenamefont {Guerin}, \citenamefont {Habegger},
  \citenamefont {Harrigan}, \citenamefont {Hartmann}, \citenamefont {Ho},
  \citenamefont {Hoffmann}, \citenamefont {Huang}, \citenamefont {Humble},
  \citenamefont {Isakov}, \citenamefont {Jeffrey}, \citenamefont {Jiang},
  \citenamefont {Kafri}, \citenamefont {Kechedzhi}, \citenamefont {Kelly},
  \citenamefont {Klimov}, \citenamefont {Knysh}, \citenamefont {Korotkov},
  \citenamefont {Kostritsa}, \citenamefont {Landhuis}, \citenamefont
  {Lindmark}, \citenamefont {Lucero}, \citenamefont {Lyakh}, \citenamefont
  {Mandrà}, \citenamefont {McClean}, \citenamefont {McEwen}, \citenamefont
  {Megrant}, \citenamefont {Mi}, \citenamefont {Michielsen}, \citenamefont
  {Mohseni}, \citenamefont {Mutus}, \citenamefont {Naaman}, \citenamefont
  {Neeley}, \citenamefont {Neill}, \citenamefont {Niu}, \citenamefont {Ostby},
  \citenamefont {Petukhov}, \citenamefont {Platt}, \citenamefont {Quintana},
  \citenamefont {Rieffel}, \citenamefont {Roushan}, \citenamefont {Rubin},
  \citenamefont {Sank}, \citenamefont {Satzinger}, \citenamefont {Smelyanskiy},
  \citenamefont {Sung}, \citenamefont {Trevithick}, \citenamefont
  {Vainsencher}, \citenamefont {Villalonga}, \citenamefont {White},
  \citenamefont {Yao}, \citenamefont {Yeh}, \citenamefont {Zalcman},
  \citenamefont {Neven},\ and\ \citenamefont {Martinis}}]{arute_quantum_2019}%
  \BibitemOpen
  \bibfield  {author} {\bibinfo {author} {\bibfnamefont {F.}~\bibnamefont
  {Arute}}, \bibinfo {author} {\bibfnamefont {K.}~\bibnamefont {Arya}},
  \bibinfo {author} {\bibfnamefont {R.}~\bibnamefont {Babbush}}, \bibinfo
  {author} {\bibfnamefont {D.}~\bibnamefont {Bacon}}, \bibinfo {author}
  {\bibfnamefont {J.~C.}\ \bibnamefont {Bardin}}, \bibinfo {author}
  {\bibfnamefont {R.}~\bibnamefont {Barends}}, \bibinfo {author} {\bibfnamefont
  {R.}~\bibnamefont {Biswas}}, \bibinfo {author} {\bibfnamefont
  {S.}~\bibnamefont {Boixo}}, \bibinfo {author} {\bibfnamefont {F.~G. S.~L.}\
  \bibnamefont {Brandao}}, \bibinfo {author} {\bibfnamefont {D.~A.}\
  \bibnamefont {Buell}}, \bibinfo {author} {\bibfnamefont {B.}~\bibnamefont
  {Burkett}}, \bibinfo {author} {\bibfnamefont {Y.}~\bibnamefont {Chen}},
  \bibinfo {author} {\bibfnamefont {Z.}~\bibnamefont {Chen}}, \bibinfo {author}
  {\bibfnamefont {B.}~\bibnamefont {Chiaro}}, \bibinfo {author} {\bibfnamefont
  {R.}~\bibnamefont {Collins}}, \bibinfo {author} {\bibfnamefont
  {W.}~\bibnamefont {Courtney}}, \bibinfo {author} {\bibfnamefont
  {A.}~\bibnamefont {Dunsworth}}, \bibinfo {author} {\bibfnamefont
  {E.}~\bibnamefont {Farhi}}, \bibinfo {author} {\bibfnamefont
  {B.}~\bibnamefont {Foxen}}, \bibinfo {author} {\bibfnamefont
  {A.}~\bibnamefont {Fowler}}, \bibinfo {author} {\bibfnamefont
  {C.}~\bibnamefont {Gidney}}, \bibinfo {author} {\bibfnamefont
  {M.}~\bibnamefont {Giustina}}, \bibinfo {author} {\bibfnamefont
  {R.}~\bibnamefont {Graff}}, \bibinfo {author} {\bibfnamefont
  {K.}~\bibnamefont {Guerin}}, \bibinfo {author} {\bibfnamefont
  {S.}~\bibnamefont {Habegger}}, \bibinfo {author} {\bibfnamefont {M.~P.}\
  \bibnamefont {Harrigan}}, \bibinfo {author} {\bibfnamefont {M.~J.}\
  \bibnamefont {Hartmann}}, \bibinfo {author} {\bibfnamefont {A.}~\bibnamefont
  {Ho}}, \bibinfo {author} {\bibfnamefont {M.}~\bibnamefont {Hoffmann}},
  \bibinfo {author} {\bibfnamefont {T.}~\bibnamefont {Huang}}, \bibinfo
  {author} {\bibfnamefont {T.~S.}\ \bibnamefont {Humble}}, \bibinfo {author}
  {\bibfnamefont {S.~V.}\ \bibnamefont {Isakov}}, \bibinfo {author}
  {\bibfnamefont {E.}~\bibnamefont {Jeffrey}}, \bibinfo {author} {\bibfnamefont
  {Z.}~\bibnamefont {Jiang}}, \bibinfo {author} {\bibfnamefont
  {D.}~\bibnamefont {Kafri}}, \bibinfo {author} {\bibfnamefont
  {K.}~\bibnamefont {Kechedzhi}}, \bibinfo {author} {\bibfnamefont
  {J.}~\bibnamefont {Kelly}}, \bibinfo {author} {\bibfnamefont {P.~V.}\
  \bibnamefont {Klimov}}, \bibinfo {author} {\bibfnamefont {S.}~\bibnamefont
  {Knysh}}, \bibinfo {author} {\bibfnamefont {A.}~\bibnamefont {Korotkov}},
  \bibinfo {author} {\bibfnamefont {F.}~\bibnamefont {Kostritsa}}, \bibinfo
  {author} {\bibfnamefont {D.}~\bibnamefont {Landhuis}}, \bibinfo {author}
  {\bibfnamefont {M.}~\bibnamefont {Lindmark}}, \bibinfo {author}
  {\bibfnamefont {E.}~\bibnamefont {Lucero}}, \bibinfo {author} {\bibfnamefont
  {D.}~\bibnamefont {Lyakh}}, \bibinfo {author} {\bibfnamefont
  {S.}~\bibnamefont {Mandrà}}, \bibinfo {author} {\bibfnamefont {J.~R.}\
  \bibnamefont {McClean}}, \bibinfo {author} {\bibfnamefont {M.}~\bibnamefont
  {McEwen}}, \bibinfo {author} {\bibfnamefont {A.}~\bibnamefont {Megrant}},
  \bibinfo {author} {\bibfnamefont {X.}~\bibnamefont {Mi}}, \bibinfo {author}
  {\bibfnamefont {K.}~\bibnamefont {Michielsen}}, \bibinfo {author}
  {\bibfnamefont {M.}~\bibnamefont {Mohseni}}, \bibinfo {author} {\bibfnamefont
  {J.}~\bibnamefont {Mutus}}, \bibinfo {author} {\bibfnamefont
  {O.}~\bibnamefont {Naaman}}, \bibinfo {author} {\bibfnamefont
  {M.}~\bibnamefont {Neeley}}, \bibinfo {author} {\bibfnamefont
  {C.}~\bibnamefont {Neill}}, \bibinfo {author} {\bibfnamefont {M.~Y.}\
  \bibnamefont {Niu}}, \bibinfo {author} {\bibfnamefont {E.}~\bibnamefont
  {Ostby}}, \bibinfo {author} {\bibfnamefont {A.}~\bibnamefont {Petukhov}},
  \bibinfo {author} {\bibfnamefont {J.~C.}\ \bibnamefont {Platt}}, \bibinfo
  {author} {\bibfnamefont {C.}~\bibnamefont {Quintana}}, \bibinfo {author}
  {\bibfnamefont {E.~G.}\ \bibnamefont {Rieffel}}, \bibinfo {author}
  {\bibfnamefont {P.}~\bibnamefont {Roushan}}, \bibinfo {author} {\bibfnamefont
  {N.~C.}\ \bibnamefont {Rubin}}, \bibinfo {author} {\bibfnamefont
  {D.}~\bibnamefont {Sank}}, \bibinfo {author} {\bibfnamefont {K.~J.}\
  \bibnamefont {Satzinger}}, \bibinfo {author} {\bibfnamefont {V.}~\bibnamefont
  {Smelyanskiy}}, \bibinfo {author} {\bibfnamefont {K.~J.}\ \bibnamefont
  {Sung}}, \bibinfo {author} {\bibfnamefont {M.~D.}\ \bibnamefont
  {Trevithick}}, \bibinfo {author} {\bibfnamefont {A.}~\bibnamefont
  {Vainsencher}}, \bibinfo {author} {\bibfnamefont {B.}~\bibnamefont
  {Villalonga}}, \bibinfo {author} {\bibfnamefont {T.}~\bibnamefont {White}},
  \bibinfo {author} {\bibfnamefont {Z.~J.}\ \bibnamefont {Yao}}, \bibinfo
  {author} {\bibfnamefont {P.}~\bibnamefont {Yeh}}, \bibinfo {author}
  {\bibfnamefont {A.}~\bibnamefont {Zalcman}}, \bibinfo {author} {\bibfnamefont
  {H.}~\bibnamefont {Neven}},\ and\ \bibinfo {author} {\bibfnamefont {J.~M.}\
  \bibnamefont {Martinis}},\ }\bibfield  {title} {\bibinfo {title} {Quantum
  supremacy using a programmable superconducting processor},\ }\href@noop {}
  {\bibfield  {journal} {\bibinfo  {journal} {Nature}\ }\textbf {\bibinfo
  {volume} {574}},\ \bibinfo {pages} {505} (\bibinfo {year}
  {2019})}\BibitemShut {NoStop}%
\bibitem [{\citenamefont {Phillips}(1972)}]{phillips_tunneling_1972}%
  \BibitemOpen
  \bibfield  {author} {\bibinfo {author} {\bibfnamefont {W.~A.}\ \bibnamefont
  {Phillips}},\ }\bibfield  {title} {\bibinfo {title} {Tunneling states in
  amorphous solids},\ }\href@noop {} {\bibfield  {journal} {\bibinfo  {journal}
  {J. Low Temp. Phys.}\ }\textbf {\bibinfo {volume} {7}},\ \bibinfo {pages}
  {351} (\bibinfo {year} {1972})}\BibitemShut {NoStop}%
\bibitem [{\citenamefont {Müller}\ \emph {et~al.}(2019)\citenamefont
  {Müller}, \citenamefont {Cole},\ and\ \citenamefont
  {Lisenfeld}}]{muller_towards_2019}%
  \BibitemOpen
  \bibfield  {author} {\bibinfo {author} {\bibfnamefont {C.}~\bibnamefont
  {Müller}}, \bibinfo {author} {\bibfnamefont {J.~H.}\ \bibnamefont {Cole}},\
  and\ \bibinfo {author} {\bibfnamefont {J.}~\bibnamefont {Lisenfeld}},\
  }\bibfield  {title} {\bibinfo {title} {Towards understanding
  two-level-systems in amorphous solids: insights from quantum circuits},\
  }\href@noop {} {\bibfield  {journal} {\bibinfo  {journal} {Rep. Prog. Phys.}\
  }\textbf {\bibinfo {volume} {82}},\ \bibinfo {pages} {124501} (\bibinfo
  {year} {2019})}\BibitemShut {NoStop}%
\bibitem [{\citenamefont {Hailey-Dunsheath}\ \emph {et~al.}(2014)\citenamefont
  {Hailey-Dunsheath}, \citenamefont {Barry}, \citenamefont {Bradford},
  \citenamefont {Chattopadhyay}, \citenamefont {Day}, \citenamefont {Doyle},
  \citenamefont {Hollister}, \citenamefont {Kovacs}, \citenamefont {LeDuc},
  \citenamefont {Llombart}, \citenamefont {Mauskopf}, \citenamefont {McKenney},
  \citenamefont {Monroe}, \citenamefont {Nguyen}, \citenamefont {O’Brient},
  \citenamefont {Padin}, \citenamefont {Reck}, \citenamefont {Shirokoff},
  \citenamefont {Swenson}, \citenamefont {Tucker},\ and\ \citenamefont
  {Zmuidzinas}}]{hailey-dunsheath_optical_2014}%
  \BibitemOpen
  \bibfield  {author} {\bibinfo {author} {\bibfnamefont {S.}~\bibnamefont
  {Hailey-Dunsheath}}, \bibinfo {author} {\bibfnamefont {P.~S.}\ \bibnamefont
  {Barry}}, \bibinfo {author} {\bibfnamefont {C.~M.}\ \bibnamefont {Bradford}},
  \bibinfo {author} {\bibfnamefont {G.}~\bibnamefont {Chattopadhyay}}, \bibinfo
  {author} {\bibfnamefont {P.}~\bibnamefont {Day}}, \bibinfo {author}
  {\bibfnamefont {S.}~\bibnamefont {Doyle}}, \bibinfo {author} {\bibfnamefont
  {M.}~\bibnamefont {Hollister}}, \bibinfo {author} {\bibfnamefont
  {A.}~\bibnamefont {Kovacs}}, \bibinfo {author} {\bibfnamefont {H.~G.}\
  \bibnamefont {LeDuc}}, \bibinfo {author} {\bibfnamefont {N.}~\bibnamefont
  {Llombart}}, \bibinfo {author} {\bibfnamefont {P.}~\bibnamefont {Mauskopf}},
  \bibinfo {author} {\bibfnamefont {C.}~\bibnamefont {McKenney}}, \bibinfo
  {author} {\bibfnamefont {R.}~\bibnamefont {Monroe}}, \bibinfo {author}
  {\bibfnamefont {H.~T.}\ \bibnamefont {Nguyen}}, \bibinfo {author}
  {\bibfnamefont {R.}~\bibnamefont {O’Brient}}, \bibinfo {author}
  {\bibfnamefont {S.}~\bibnamefont {Padin}}, \bibinfo {author} {\bibfnamefont
  {T.}~\bibnamefont {Reck}}, \bibinfo {author} {\bibfnamefont {E.}~\bibnamefont
  {Shirokoff}}, \bibinfo {author} {\bibfnamefont {L.}~\bibnamefont {Swenson}},
  \bibinfo {author} {\bibfnamefont {C.~E.}\ \bibnamefont {Tucker}},\ and\
  \bibinfo {author} {\bibfnamefont {J.}~\bibnamefont {Zmuidzinas}},\ }\bibfield
   {title} {\bibinfo {title} {Optical {Measurements} of {SuperSpec}: {A}
  {Millimeter}-{Wave} {On}-{Chip} {Spectrometer}},\ }\href@noop {} {\bibfield
  {journal} {\bibinfo  {journal} {J. Low Temp. Phys.}\ }\textbf {\bibinfo
  {volume} {176}},\ \bibinfo {pages} {841} (\bibinfo {year}
  {2014})}\BibitemShut {NoStop}%
\bibitem [{\citenamefont {Valenzuela}\ \emph {et~al.}(2019)\citenamefont
  {Valenzuela}, \citenamefont {Mena},\ and\ \citenamefont
  {Baselmans}}]{valenzuela_modelling_2019}%
  \BibitemOpen
  \bibfield  {author} {\bibinfo {author} {\bibfnamefont {D.}~\bibnamefont
  {Valenzuela}}, \bibinfo {author} {\bibfnamefont {F.~P.}\ \bibnamefont
  {Mena}},\ and\ \bibinfo {author} {\bibfnamefont {J.}~\bibnamefont
  {Baselmans}},\ }\bibfield  {title} {\bibinfo {title} {Modelling dielectric
  losses in microstrip traveling-wave kinetic-inductance parametric
  amplifiers},\ }in\ \href@noop {} {\emph {\bibinfo {booktitle} {30th
  {International} {Symposium} on {Space} {THz} {Technology}}}}\ (\bibinfo
  {address} {Gothenburg, Sweden},\ \bibinfo {year} {2019})\ pp.\ \bibinfo
  {pages} {63--66}\BibitemShut {NoStop}%
\bibitem [{\citenamefont {Shan}\ \emph {et~al.}(2016)\citenamefont {Shan},
  \citenamefont {Sekimoto},\ and\ \citenamefont
  {Noguchi}}]{shan_parametric_2016}%
  \BibitemOpen
  \bibfield  {author} {\bibinfo {author} {\bibfnamefont {W.}~\bibnamefont
  {Shan}}, \bibinfo {author} {\bibfnamefont {Y.}~\bibnamefont {Sekimoto}},\
  and\ \bibinfo {author} {\bibfnamefont {T.}~\bibnamefont {Noguchi}},\
  }\bibfield  {title} {\bibinfo {title} {Parametric {Amplification} in a
  {Superconducting} {Microstrip} {Transmission} {Line}},\ }\href@noop {}
  {\bibfield  {journal} {\bibinfo  {journal} {IEEE Trans. Appl. Supercond.}\
  }\textbf {\bibinfo {volume} {26}},\ \bibinfo {pages} {1} (\bibinfo {year}
  {2016})}\BibitemShut {NoStop}%
\bibitem [{\citenamefont {Martinis}\ \emph {et~al.}(2005)\citenamefont
  {Martinis}, \citenamefont {Cooper}, \citenamefont {McDermott}, \citenamefont
  {Steffen}, \citenamefont {Ansmann}, \citenamefont {Osborn}, \citenamefont
  {Cicak}, \citenamefont {Oh}, \citenamefont {Pappas}, \citenamefont
  {Simmonds},\ and\ \citenamefont {Yu}}]{martinis_decoherence_2005}%
  \BibitemOpen
  \bibfield  {author} {\bibinfo {author} {\bibfnamefont {J.~M.}\ \bibnamefont
  {Martinis}}, \bibinfo {author} {\bibfnamefont {K.~B.}\ \bibnamefont
  {Cooper}}, \bibinfo {author} {\bibfnamefont {R.}~\bibnamefont {McDermott}},
  \bibinfo {author} {\bibfnamefont {M.}~\bibnamefont {Steffen}}, \bibinfo
  {author} {\bibfnamefont {M.}~\bibnamefont {Ansmann}}, \bibinfo {author}
  {\bibfnamefont {K.~D.}\ \bibnamefont {Osborn}}, \bibinfo {author}
  {\bibfnamefont {K.}~\bibnamefont {Cicak}}, \bibinfo {author} {\bibfnamefont
  {S.}~\bibnamefont {Oh}}, \bibinfo {author} {\bibfnamefont {D.~P.}\
  \bibnamefont {Pappas}}, \bibinfo {author} {\bibfnamefont {R.~W.}\
  \bibnamefont {Simmonds}},\ and\ \bibinfo {author} {\bibfnamefont {C.~C.}\
  \bibnamefont {Yu}},\ }\bibfield  {title} {\bibinfo {title} {Decoherence in
  {Josephson} {Qubits} from {Dielectric} {Loss}},\ }\href@noop {} {\bibfield
  {journal} {\bibinfo  {journal} {Phys. Rev. Lett.}\ }\textbf {\bibinfo
  {volume} {95}},\ \bibinfo {pages} {210503} (\bibinfo {year}
  {2005})}\BibitemShut {NoStop}%
\bibitem [{\citenamefont {Hähnle}\ \emph {et~al.}(2020)\citenamefont
  {Hähnle}, \citenamefont {Marrewijk}, \citenamefont {Endo}, \citenamefont
  {Karatsu}, \citenamefont {Thoen}, \citenamefont {Murugesan},\ and\
  \citenamefont {Baselmans}}]{hahnle_suppression_2020}%
  \BibitemOpen
  \bibfield  {author} {\bibinfo {author} {\bibfnamefont {S.}~\bibnamefont
  {Hähnle}}, \bibinfo {author} {\bibfnamefont {N.~v.}\ \bibnamefont
  {Marrewijk}}, \bibinfo {author} {\bibfnamefont {A.}~\bibnamefont {Endo}},
  \bibinfo {author} {\bibfnamefont {K.}~\bibnamefont {Karatsu}}, \bibinfo
  {author} {\bibfnamefont {D.~J.}\ \bibnamefont {Thoen}}, \bibinfo {author}
  {\bibfnamefont {V.}~\bibnamefont {Murugesan}},\ and\ \bibinfo {author}
  {\bibfnamefont {J.~J.~A.}\ \bibnamefont {Baselmans}},\ }\bibfield  {title}
  {\bibinfo {title} {Suppression of radiation loss in high kinetic inductance
  superconducting co-planar waveguides},\ }\href@noop {} {\bibfield  {journal}
  {\bibinfo  {journal} {Appl. Phys. Lett.}\ }\textbf {\bibinfo {volume}
  {116}},\ \bibinfo {pages} {182601} (\bibinfo {year} {2020})}\BibitemShut
  {NoStop}%
\bibitem [{\citenamefont {Mazin}\ \emph {et~al.}(2010)\citenamefont {Mazin},
  \citenamefont {Sank}, \citenamefont {McHugh}, \citenamefont {Lucero},
  \citenamefont {Merrill}, \citenamefont {Gao}, \citenamefont {Pappas},
  \citenamefont {Moore},\ and\ \citenamefont {Zmuidzinas}}]{mazin_thin_2010}%
  \BibitemOpen
  \bibfield  {author} {\bibinfo {author} {\bibfnamefont {B.~A.}\ \bibnamefont
  {Mazin}}, \bibinfo {author} {\bibfnamefont {D.}~\bibnamefont {Sank}},
  \bibinfo {author} {\bibfnamefont {S.}~\bibnamefont {McHugh}}, \bibinfo
  {author} {\bibfnamefont {E.~A.}\ \bibnamefont {Lucero}}, \bibinfo {author}
  {\bibfnamefont {A.}~\bibnamefont {Merrill}}, \bibinfo {author} {\bibfnamefont
  {J.}~\bibnamefont {Gao}}, \bibinfo {author} {\bibfnamefont {D.}~\bibnamefont
  {Pappas}}, \bibinfo {author} {\bibfnamefont {D.}~\bibnamefont {Moore}},\ and\
  \bibinfo {author} {\bibfnamefont {J.}~\bibnamefont {Zmuidzinas}},\ }\bibfield
   {title} {\bibinfo {title} {Thin film dielectric microstrip kinetic
  inductance detectors},\ }\href@noop {} {\bibfield  {journal} {\bibinfo
  {journal} {Appl. Phys. Lett.}\ }\textbf {\bibinfo {volume} {96}},\ \bibinfo
  {pages} {102504} (\bibinfo {year} {2010})}\BibitemShut {NoStop}%
\bibitem [{\citenamefont {O’Connell}\ \emph {et~al.}(2008)\citenamefont
  {O’Connell}, \citenamefont {Ansmann}, \citenamefont {Bialczak},
  \citenamefont {Hofheinz}, \citenamefont {Katz}, \citenamefont {Lucero},
  \citenamefont {McKenney}, \citenamefont {Neeley}, \citenamefont {Wang},
  \citenamefont {Weig}, \citenamefont {Cleland},\ and\ \citenamefont
  {Martinis}}]{oconnell_microwave_2008}%
  \BibitemOpen
  \bibfield  {author} {\bibinfo {author} {\bibfnamefont {A.~D.}\ \bibnamefont
  {O’Connell}}, \bibinfo {author} {\bibfnamefont {M.}~\bibnamefont
  {Ansmann}}, \bibinfo {author} {\bibfnamefont {R.~C.}\ \bibnamefont
  {Bialczak}}, \bibinfo {author} {\bibfnamefont {M.}~\bibnamefont {Hofheinz}},
  \bibinfo {author} {\bibfnamefont {N.}~\bibnamefont {Katz}}, \bibinfo {author}
  {\bibfnamefont {E.}~\bibnamefont {Lucero}}, \bibinfo {author} {\bibfnamefont
  {C.}~\bibnamefont {McKenney}}, \bibinfo {author} {\bibfnamefont
  {M.}~\bibnamefont {Neeley}}, \bibinfo {author} {\bibfnamefont
  {H.}~\bibnamefont {Wang}}, \bibinfo {author} {\bibfnamefont {E.~M.}\
  \bibnamefont {Weig}}, \bibinfo {author} {\bibfnamefont {A.~N.}\ \bibnamefont
  {Cleland}},\ and\ \bibinfo {author} {\bibfnamefont {J.~M.}\ \bibnamefont
  {Martinis}},\ }\bibfield  {title} {\bibinfo {title} {Microwave dielectric
  loss at single photon energies and millikelvin temperatures},\ }\href@noop {}
  {\bibfield  {journal} {\bibinfo  {journal} {Appl. Phys. Lett.}\ }\textbf
  {\bibinfo {volume} {92}},\ \bibinfo {pages} {112903} (\bibinfo {year}
  {2008})}\BibitemShut {NoStop}%
\bibitem [{\citenamefont {Hähnle}\ \emph {et~al.}(2021)\citenamefont
  {Hähnle}, \citenamefont {Kouwenhoven}, \citenamefont {Buijtendorp},
  \citenamefont {Endo}, \citenamefont {Karatsu}, \citenamefont {Thoen},
  \citenamefont {Murugesan},\ and\ \citenamefont
  {Baselmans}}]{hahnle_superconducting_2021}%
  \BibitemOpen
  \bibfield  {author} {\bibinfo {author} {\bibfnamefont {S.}~\bibnamefont
  {Hähnle}}, \bibinfo {author} {\bibfnamefont {K.}~\bibnamefont
  {Kouwenhoven}}, \bibinfo {author} {\bibfnamefont {B.}~\bibnamefont
  {Buijtendorp}}, \bibinfo {author} {\bibfnamefont {A.}~\bibnamefont {Endo}},
  \bibinfo {author} {\bibfnamefont {K.}~\bibnamefont {Karatsu}}, \bibinfo
  {author} {\bibfnamefont {D.}~\bibnamefont {Thoen}}, \bibinfo {author}
  {\bibfnamefont {V.}~\bibnamefont {Murugesan}},\ and\ \bibinfo {author}
  {\bibfnamefont {J.}~\bibnamefont {Baselmans}},\ }\bibfield  {title} {\bibinfo
  {title} {Superconducting {Microstrip} {Losses} at {Microwave} and
  {Submillimeter} {Wavelengths}},\ }\href@noop {} {\bibfield  {journal}
  {\bibinfo  {journal} {Phys. Rev. Appl.}\ }\textbf {\bibinfo {volume} {16}},\
  \bibinfo {pages} {014019} (\bibinfo {year} {2021})}\BibitemShut {NoStop}%
\bibitem [{\citenamefont {Buijtendorp}\ \emph {et~al.}(2020)\citenamefont
  {Buijtendorp}, \citenamefont {Bueno}, \citenamefont {Thoen}, \citenamefont
  {Murugesan}, \citenamefont {Sberna}, \citenamefont {Baselmans}, \citenamefont
  {Vollebregt},\ and\ \citenamefont
  {Endo}}]{buijtendorp_characterization_2020}%
  \BibitemOpen
  \bibfield  {author} {\bibinfo {author} {\bibfnamefont {B.}~\bibnamefont
  {Buijtendorp}}, \bibinfo {author} {\bibfnamefont {J.}~\bibnamefont {Bueno}},
  \bibinfo {author} {\bibfnamefont {D.~J.}\ \bibnamefont {Thoen}}, \bibinfo
  {author} {\bibfnamefont {V.}~\bibnamefont {Murugesan}}, \bibinfo {author}
  {\bibfnamefont {P.}~\bibnamefont {Sberna}}, \bibinfo {author} {\bibfnamefont
  {J.~J.~A.}\ \bibnamefont {Baselmans}}, \bibinfo {author} {\bibfnamefont
  {S.}~\bibnamefont {Vollebregt}},\ and\ \bibinfo {author} {\bibfnamefont
  {A.}~\bibnamefont {Endo}},\ }\bibfield  {title} {\bibinfo {title}
  {Characterization of low-loss hydrogenated amorphous silicon ﬁlms for
  superconducting resonators},\ }in\ \href@noop {} {\emph {\bibinfo {booktitle}
  {Millimeter, {Submillimeter}, and {Far}-{Infrared} {Detectors} and
  {Instrumentation} for {Astronomy} {X}}}},\ Vol.\ \bibinfo {volume} {11453
  114532X-1},\ \bibinfo {editor} {edited by\ \bibinfo {editor} {\bibfnamefont
  {J.}~\bibnamefont {Zmuidzinas}}\ and\ \bibinfo {editor} {\bibfnamefont
  {J.-R.}\ \bibnamefont {Gao}}}\ (\bibinfo  {publisher} {SPIE},\ \bibinfo
  {year} {2020})\BibitemShut {NoStop}%
\bibitem [{\citenamefont {Gao}\ \emph {et~al.}(2009)\citenamefont {Gao},
  \citenamefont {Vayonakis}, \citenamefont {Noroozian}, \citenamefont
  {Zmuidzinas}, \citenamefont {Day}, \citenamefont {Leduc}, \citenamefont
  {Young}, \citenamefont {Cabrera},\ and\ \citenamefont
  {Miller}}]{gao_measurement_2009}%
  \BibitemOpen
  \bibfield  {author} {\bibinfo {author} {\bibfnamefont {J.}~\bibnamefont
  {Gao}}, \bibinfo {author} {\bibfnamefont {A.}~\bibnamefont {Vayonakis}},
  \bibinfo {author} {\bibfnamefont {O.}~\bibnamefont {Noroozian}}, \bibinfo
  {author} {\bibfnamefont {J.}~\bibnamefont {Zmuidzinas}}, \bibinfo {author}
  {\bibfnamefont {P.~K.}\ \bibnamefont {Day}}, \bibinfo {author} {\bibfnamefont
  {H.~G.}\ \bibnamefont {Leduc}}, \bibinfo {author} {\bibfnamefont
  {B.}~\bibnamefont {Young}}, \bibinfo {author} {\bibfnamefont
  {B.}~\bibnamefont {Cabrera}},\ and\ \bibinfo {author} {\bibfnamefont
  {A.}~\bibnamefont {Miller}},\ }\bibfield  {title} {\bibinfo {title}
  {Measurement of loss in superconducting microstrip at millimeter-wave
  frequencies}\ }(\bibinfo {address} {Stanford (California)},\ \bibinfo {year}
  {2009})\ pp.\ \bibinfo {pages} {164--167}\BibitemShut {NoStop}%
\bibitem [{\citenamefont {Endo}\ \emph {et~al.}(2013)\citenamefont {Endo},
  \citenamefont {Sfiligoj}, \citenamefont {Yates}, \citenamefont {Baselmans},
  \citenamefont {Thoen}, \citenamefont {Javadzadeh}, \citenamefont {van~der
  Werf}, \citenamefont {Baryshev},\ and\ \citenamefont
  {Klapwijk}}]{endo_-chip_2013}%
  \BibitemOpen
  \bibfield  {author} {\bibinfo {author} {\bibfnamefont {A.}~\bibnamefont
  {Endo}}, \bibinfo {author} {\bibfnamefont {C.}~\bibnamefont {Sfiligoj}},
  \bibinfo {author} {\bibfnamefont {S.~J.~C.}\ \bibnamefont {Yates}}, \bibinfo
  {author} {\bibfnamefont {J.~J.~A.}\ \bibnamefont {Baselmans}}, \bibinfo
  {author} {\bibfnamefont {D.~J.}\ \bibnamefont {Thoen}}, \bibinfo {author}
  {\bibfnamefont {S.~M.~H.}\ \bibnamefont {Javadzadeh}}, \bibinfo {author}
  {\bibfnamefont {P.~P.}\ \bibnamefont {van~der Werf}}, \bibinfo {author}
  {\bibfnamefont {A.~M.}\ \bibnamefont {Baryshev}},\ and\ \bibinfo {author}
  {\bibfnamefont {T.~M.}\ \bibnamefont {Klapwijk}},\ }\bibfield  {title}
  {\bibinfo {title} {On-chip filter bank spectroscopy at 600–700 {GHz} using
  {NbTiN} superconducting resonators},\ }\href@noop {} {\bibfield  {journal}
  {\bibinfo  {journal} {Appl. Phys. Lett.}\ }\textbf {\bibinfo {volume}
  {103}},\ \bibinfo {pages} {032601} (\bibinfo {year} {2013})}\BibitemShut
  {NoStop}%
\bibitem [{\citenamefont {Molina-Ruiz}\ \emph {et~al.}(2021)\citenamefont
  {Molina-Ruiz}, \citenamefont {Rosen}, \citenamefont {Jacks}, \citenamefont
  {Abernathy}, \citenamefont {Metcalf}, \citenamefont {Liu}, \citenamefont
  {DuBois},\ and\ \citenamefont {Hellman}}]{molina-ruiz_origin_2021}%
  \BibitemOpen
  \bibfield  {author} {\bibinfo {author} {\bibfnamefont {M.}~\bibnamefont
  {Molina-Ruiz}}, \bibinfo {author} {\bibfnamefont {Y.~J.}\ \bibnamefont
  {Rosen}}, \bibinfo {author} {\bibfnamefont {H.~C.}\ \bibnamefont {Jacks}},
  \bibinfo {author} {\bibfnamefont {M.~R.}\ \bibnamefont {Abernathy}}, \bibinfo
  {author} {\bibfnamefont {T.~H.}\ \bibnamefont {Metcalf}}, \bibinfo {author}
  {\bibfnamefont {X.}~\bibnamefont {Liu}}, \bibinfo {author} {\bibfnamefont
  {J.~L.}\ \bibnamefont {DuBois}},\ and\ \bibinfo {author} {\bibfnamefont
  {F.}~\bibnamefont {Hellman}},\ }\bibfield  {title} {\bibinfo {title} {Origin
  of mechanical and dielectric losses from two-level systems in amorphous
  silicon},\ }\href@noop {} {\bibfield  {journal} {\bibinfo  {journal} {Phys.
  Rev. Mater.}\ }\textbf {\bibinfo {volume} {5}},\ \bibinfo {pages} {035601}
  (\bibinfo {year} {2021})}\BibitemShut {NoStop}%
\bibitem [{\citenamefont {Sarro}\ \emph {et~al.}(1998)\citenamefont {Sarro},
  \citenamefont {de~Boer}, \citenamefont {Korkmaz},\ and\ \citenamefont
  {Laros}}]{sarro_low-stress_1998}%
  \BibitemOpen
  \bibfield  {author} {\bibinfo {author} {\bibfnamefont {P.}~\bibnamefont
  {Sarro}}, \bibinfo {author} {\bibfnamefont {C.}~\bibnamefont {de~Boer}},
  \bibinfo {author} {\bibfnamefont {E.}~\bibnamefont {Korkmaz}},\ and\ \bibinfo
  {author} {\bibfnamefont {J.}~\bibnamefont {Laros}},\ }\bibfield  {title}
  {\bibinfo {title} {Low-stress {PECVD} {SiC} thin films for {IC}-compatible
  microstructures},\ }\href@noop {} {\bibfield  {journal} {\bibinfo  {journal}
  {Sens. Actuator A Phys.}\ }\textbf {\bibinfo {volume} {67}},\ \bibinfo
  {pages} {175} (\bibinfo {year} {1998})}\BibitemShut {NoStop}%
\bibitem [{\citenamefont {de~Visser}\ \emph {et~al.}(2021)\citenamefont
  {de~Visser}, \citenamefont {de~Rooij}, \citenamefont {Murugesan},
  \citenamefont {Thoen},\ and\ \citenamefont
  {Baselmans}}]{de_visser_phonon-trapping_2021}%
  \BibitemOpen
  \bibfield  {author} {\bibinfo {author} {\bibfnamefont {P.~J.}\ \bibnamefont
  {de~Visser}}, \bibinfo {author} {\bibfnamefont {S.~A.~H.}\ \bibnamefont
  {de~Rooij}}, \bibinfo {author} {\bibfnamefont {V.}~\bibnamefont {Murugesan}},
  \bibinfo {author} {\bibfnamefont {D.~J.}\ \bibnamefont {Thoen}},\ and\
  \bibinfo {author} {\bibfnamefont {J.~J.~A.}\ \bibnamefont {Baselmans}},\
  }\bibfield  {title} {\bibinfo {title} {Phonon-trapping enhanced energy
  resolution in superconducting single photon detectors},\ }\href
  {http://arxiv.org/abs/2103.06723} {\bibfield  {journal} {\bibinfo  {journal}
  {preprint, arXiv:2103.06723 [astro-ph, physics:cond-mat, physics:physics]}\ }
  (\bibinfo {year} {2021})}\BibitemShut {NoStop}%
\bibitem [{\citenamefont {Mishima}\ and\ \citenamefont
  {Yagishita}(1988)}]{mishima_investigation_1988}%
  \BibitemOpen
  \bibfield  {author} {\bibinfo {author} {\bibfnamefont {Y.}~\bibnamefont
  {Mishima}}\ and\ \bibinfo {author} {\bibfnamefont {T.}~\bibnamefont
  {Yagishita}},\ }\bibfield  {title} {\bibinfo {title} {Investigation of the
  bubble formation mechanism in a‐{Si}:{H} films by {Fourier}‐transform
  infrared microspectroscopy},\ }\href@noop {} {\bibfield  {journal} {\bibinfo
  {journal} {J. Appl. Phys.}\ }\textbf {\bibinfo {volume} {64}},\ \bibinfo
  {pages} {3972} (\bibinfo {year} {1988})}\BibitemShut {NoStop}%
\bibitem [{\citenamefont {Wang}\ \emph {et~al.}(2018)\citenamefont {Wang},
  \citenamefont {Wu}, \citenamefont {Chen}, \citenamefont {Zhuo},\ and\
  \citenamefont {Wang}}]{wang_avoiding_2018}%
  \BibitemOpen
  \bibfield  {author} {\bibinfo {author} {\bibfnamefont {J.}~\bibnamefont
  {Wang}}, \bibinfo {author} {\bibfnamefont {L.}~\bibnamefont {Wu}}, \bibinfo
  {author} {\bibfnamefont {X.}~\bibnamefont {Chen}}, \bibinfo {author}
  {\bibfnamefont {W.}~\bibnamefont {Zhuo}},\ and\ \bibinfo {author}
  {\bibfnamefont {G.}~\bibnamefont {Wang}},\ }\bibfield  {title} {\bibinfo
  {title} {Avoiding blister defects in low-stress hydrogenated amorphous
  silicon films for {MEMS} sensors},\ }\href@noop {} {\bibfield  {journal}
  {\bibinfo  {journal} {Sens. Actuator A Phys.}\ }\textbf {\bibinfo {volume}
  {276}},\ \bibinfo {pages} {11} (\bibinfo {year} {2018})}\BibitemShut
  {NoStop}%
\bibitem [{\citenamefont {van~de Ven}(1988)}]{van_de_ven_advances_1988}%
  \BibitemOpen
  \bibfield  {author} {\bibinfo {author} {\bibfnamefont {E.~P.}\ \bibnamefont
  {van~de Ven}},\ }\bibfield  {title} {\bibinfo {title} {Advances {In} {Plasma}
  {Enhanced} {Thin} {Film} {Deposition}}\ }(\bibinfo {address} {Santa Clara,
  CA},\ \bibinfo {year} {1988})\ p.\ \bibinfo {pages} {117}\BibitemShut
  {NoStop}%
\bibitem [{\citenamefont {Janssen}\ \emph {et~al.}(2013)\citenamefont
  {Janssen}, \citenamefont {Baselmans}, \citenamefont {Endo}, \citenamefont
  {Ferrari}, \citenamefont {Yates}, \citenamefont {Baryshev},\ and\
  \citenamefont {Klapwijk}}]{janssen_high_2013}%
  \BibitemOpen
  \bibfield  {author} {\bibinfo {author} {\bibfnamefont {R.~M.~J.}\
  \bibnamefont {Janssen}}, \bibinfo {author} {\bibfnamefont {J.~J.~A.}\
  \bibnamefont {Baselmans}}, \bibinfo {author} {\bibfnamefont {A.}~\bibnamefont
  {Endo}}, \bibinfo {author} {\bibfnamefont {L.}~\bibnamefont {Ferrari}},
  \bibinfo {author} {\bibfnamefont {S.~J.~C.}\ \bibnamefont {Yates}}, \bibinfo
  {author} {\bibfnamefont {A.~M.}\ \bibnamefont {Baryshev}},\ and\ \bibinfo
  {author} {\bibfnamefont {T.~M.}\ \bibnamefont {Klapwijk}},\ }\bibfield
  {title} {\bibinfo {title} {High optical efficiency and photon noise limited
  sensitivity of microwave kinetic inductance detectors using phase readout},\
  }\href@noop {} {\bibfield  {journal} {\bibinfo  {journal} {Appl. Phys.
  Lett.}\ }\textbf {\bibinfo {volume} {103}},\ \bibinfo {pages} {203503}
  (\bibinfo {year} {2013})}\BibitemShut {NoStop}%
\bibitem [{\citenamefont {van Rantwijk}\ \emph {et~al.}(2016)\citenamefont {van
  Rantwijk}, \citenamefont {Grim}, \citenamefont {van Loon}, \citenamefont
  {Yates}, \citenamefont {Baryshev},\ and\ \citenamefont
  {Baselmans}}]{van_rantwijk_multiplexed_2016}%
  \BibitemOpen
  \bibfield  {author} {\bibinfo {author} {\bibfnamefont {J.}~\bibnamefont {van
  Rantwijk}}, \bibinfo {author} {\bibfnamefont {M.}~\bibnamefont {Grim}},
  \bibinfo {author} {\bibfnamefont {D.}~\bibnamefont {van Loon}}, \bibinfo
  {author} {\bibfnamefont {S.}~\bibnamefont {Yates}}, \bibinfo {author}
  {\bibfnamefont {A.}~\bibnamefont {Baryshev}},\ and\ \bibinfo {author}
  {\bibfnamefont {J.}~\bibnamefont {Baselmans}},\ }\bibfield  {title} {\bibinfo
  {title} {Multiplexed {Readout} for 1000-{Pixel} {Arrays} of {Microwave}
  {Kinetic} {Inductance} {Detectors}},\ }\href@noop {} {\bibfield  {journal}
  {\bibinfo  {journal} {IEEE Trans. Microwave Theory Techn.}\ }\textbf
  {\bibinfo {volume} {64}},\ \bibinfo {pages} {1876} (\bibinfo {year}
  {2016})}\BibitemShut {NoStop}%
\bibitem [{\citenamefont {Gao}(2008)}]{gao_physics_2008}%
  \BibitemOpen
  \bibfield  {author} {\bibinfo {author} {\bibfnamefont {J.}~\bibnamefont
  {Gao}},\ }\emph {\bibinfo {title} {The {Physics} of {Superconducting}
  {Microwave} {Resonators}}},\ \href@noop {} {Ph.D. thesis},\ \bibinfo
  {school} {California Institute of Technology} (\bibinfo {year}
  {2008})\BibitemShut {NoStop}%
\bibitem [{son()}]{sonnet_user_guide}%
  \BibitemOpen
  \href@noop {} {\bibinfo {title} {{Sonnet User's Guide | Sonnet 17}}},\
  \bibinfo {howpublished}
  {\url{http://www.sonnetsoftware.com/support/manuals.asp}}\BibitemShut
  {NoStop}%
\bibitem [{\citenamefont {Khalil}\ \emph {et~al.}(2012)\citenamefont {Khalil},
  \citenamefont {Stoutimore}, \citenamefont {Wellstood},\ and\ \citenamefont
  {Osborn}}]{khalil_analysis_2012}%
  \BibitemOpen
  \bibfield  {author} {\bibinfo {author} {\bibfnamefont {M.~S.}\ \bibnamefont
  {Khalil}}, \bibinfo {author} {\bibfnamefont {M.~J.~A.}\ \bibnamefont
  {Stoutimore}}, \bibinfo {author} {\bibfnamefont {F.~C.}\ \bibnamefont
  {Wellstood}},\ and\ \bibinfo {author} {\bibfnamefont {K.~D.}\ \bibnamefont
  {Osborn}},\ }\bibfield  {title} {\bibinfo {title} {An analysis method for
  asymmetric resonator transmission applied to superconducting devices},\
  }\href@noop {} {\bibfield  {journal} {\bibinfo  {journal} {J. Appl. Phys}\
  }\textbf {\bibinfo {volume} {111}},\ \bibinfo {pages} {054510} (\bibinfo
  {year} {2012})}\BibitemShut {NoStop}%
\bibitem [{\citenamefont {Cataldo}\ \emph {et~al.}(2012)\citenamefont
  {Cataldo}, \citenamefont {Beall}, \citenamefont {Cho}, \citenamefont
  {McAndrew}, \citenamefont {Niemack},\ and\ \citenamefont
  {Wollack}}]{cataldo_infrared_2012}%
  \BibitemOpen
  \bibfield  {author} {\bibinfo {author} {\bibfnamefont {G.}~\bibnamefont
  {Cataldo}}, \bibinfo {author} {\bibfnamefont {J.~A.}\ \bibnamefont {Beall}},
  \bibinfo {author} {\bibfnamefont {H.-M.}\ \bibnamefont {Cho}}, \bibinfo
  {author} {\bibfnamefont {B.}~\bibnamefont {McAndrew}}, \bibinfo {author}
  {\bibfnamefont {M.~D.}\ \bibnamefont {Niemack}},\ and\ \bibinfo {author}
  {\bibfnamefont {E.~J.}\ \bibnamefont {Wollack}},\ }\bibfield  {title}
  {\bibinfo {title} {Infrared dielectric properties of low-stress silicon
  nitride},\ }\href@noop {} {\bibfield  {journal} {\bibinfo  {journal} {Opt.
  Lett.}\ }\textbf {\bibinfo {volume} {37}},\ \bibinfo {pages} {4200} (\bibinfo
  {year} {2012})}\BibitemShut {NoStop}%
\bibitem [{\citenamefont {Cataldo}\ \emph {et~al.}(2016)\citenamefont
  {Cataldo}, \citenamefont {Wollack}, \citenamefont {Brown},\ and\
  \citenamefont {Miller}}]{cataldo_infrared_2016}%
  \BibitemOpen
  \bibfield  {author} {\bibinfo {author} {\bibfnamefont {G.}~\bibnamefont
  {Cataldo}}, \bibinfo {author} {\bibfnamefont {E.~J.}\ \bibnamefont
  {Wollack}}, \bibinfo {author} {\bibfnamefont {A.~D.}\ \bibnamefont {Brown}},\
  and\ \bibinfo {author} {\bibfnamefont {K.~H.}\ \bibnamefont {Miller}},\
  }\bibfield  {title} {\bibinfo {title} {Infrared dielectric properties of
  low-stress silicon oxide},\ }\href@noop {} {\bibfield  {journal} {\bibinfo
  {journal} {Opt. Lett.}\ }\textbf {\bibinfo {volume} {41}},\ \bibinfo {pages}
  {1364} (\bibinfo {year} {2016})}\BibitemShut {NoStop}%
\bibitem [{\citenamefont {Faoro}\ and\ \citenamefont
  {Ioffe}(2015)}]{faoro_interacting_2015}%
  \BibitemOpen
  \bibfield  {author} {\bibinfo {author} {\bibfnamefont {L.}~\bibnamefont
  {Faoro}}\ and\ \bibinfo {author} {\bibfnamefont {L.~B.}\ \bibnamefont
  {Ioffe}},\ }\bibfield  {title} {\bibinfo {title} {Interacting tunneling model
  for two-level systems in amorphous materials and its predictions for their
  dephasing and noise in superconducting microresonators},\ }\href@noop {}
  {\bibfield  {journal} {\bibinfo  {journal} {Phys. Rev. B}\ }\textbf {\bibinfo
  {volume} {91}},\ \bibinfo {pages} {014201} (\bibinfo {year}
  {2015})}\BibitemShut {NoStop}%
\bibitem [{\citenamefont {Woods}\ \emph {et~al.}(2019)\citenamefont {Woods},
  \citenamefont {Calusine}, \citenamefont {Melville}, \citenamefont {Sevi},
  \citenamefont {Golden}, \citenamefont {Kim}, \citenamefont {Rosenberg},
  \citenamefont {Yoder},\ and\ \citenamefont
  {Oliver}}]{woods_determining_2019}%
  \BibitemOpen
  \bibfield  {author} {\bibinfo {author} {\bibfnamefont {W.}~\bibnamefont
  {Woods}}, \bibinfo {author} {\bibfnamefont {G.}~\bibnamefont {Calusine}},
  \bibinfo {author} {\bibfnamefont {A.}~\bibnamefont {Melville}}, \bibinfo
  {author} {\bibfnamefont {A.}~\bibnamefont {Sevi}}, \bibinfo {author}
  {\bibfnamefont {E.}~\bibnamefont {Golden}}, \bibinfo {author} {\bibfnamefont
  {D.}~\bibnamefont {Kim}}, \bibinfo {author} {\bibfnamefont {D.}~\bibnamefont
  {Rosenberg}}, \bibinfo {author} {\bibfnamefont {J.}~\bibnamefont {Yoder}},\
  and\ \bibinfo {author} {\bibfnamefont {W.}~\bibnamefont {Oliver}},\
  }\bibfield  {title} {\bibinfo {title} {Determining {Interface} {Dielectric}
  {Losses} in {Superconducting} {Coplanar}-{Waveguide} {Resonators}},\
  }\href@noop {} {\bibfield  {journal} {\bibinfo  {journal} {Phys. Rev. Appl.}\
  }\textbf {\bibinfo {volume} {12}},\ \bibinfo {pages} {014012} (\bibinfo
  {year} {2019})}\BibitemShut {NoStop}%
\bibitem [{\citenamefont {Gao}\ \emph {et~al.}(2008)\citenamefont {Gao},
  \citenamefont {Daal}, \citenamefont {Vayonakis}, \citenamefont {Kumar},
  \citenamefont {Zmuidzinas}, \citenamefont {Sadoulet}, \citenamefont {Mazin},
  \citenamefont {Day},\ and\ \citenamefont {Leduc}}]{gao_experimental_2008}%
  \BibitemOpen
  \bibfield  {author} {\bibinfo {author} {\bibfnamefont {J.}~\bibnamefont
  {Gao}}, \bibinfo {author} {\bibfnamefont {M.}~\bibnamefont {Daal}}, \bibinfo
  {author} {\bibfnamefont {A.}~\bibnamefont {Vayonakis}}, \bibinfo {author}
  {\bibfnamefont {S.}~\bibnamefont {Kumar}}, \bibinfo {author} {\bibfnamefont
  {J.}~\bibnamefont {Zmuidzinas}}, \bibinfo {author} {\bibfnamefont
  {B.}~\bibnamefont {Sadoulet}}, \bibinfo {author} {\bibfnamefont {B.~A.}\
  \bibnamefont {Mazin}}, \bibinfo {author} {\bibfnamefont {P.~K.}\ \bibnamefont
  {Day}},\ and\ \bibinfo {author} {\bibfnamefont {H.~G.}\ \bibnamefont
  {Leduc}},\ }\bibfield  {title} {\bibinfo {title} {Experimental evidence for a
  surface distribution of two-level systems in superconducting lithographed
  microwave resonators},\ }\href@noop {} {\bibfield  {journal} {\bibinfo
  {journal} {Appl. Phys. Lett.}\ }\textbf {\bibinfo {volume} {92}},\ \bibinfo
  {pages} {152505} (\bibinfo {year} {2008})}\BibitemShut {NoStop}%
\end{thebibliography}%

\end{document}